\documentclass{optica-article}

\journal{opticajournal}

\articletype{Research Article}

\usepackage{lineno}
\usepackage{soul}
\usepackage{makecell}
\usepackage[colorinlistoftodos]{todonotes}

\begin{document}

\title{Comparison of Lumerical FDTD and Tidy3D for three-dimensional FDTD simulations of passive silicon photonic components}

\author{Zuyang Liu\authormark{1*} and Joyce K. S. Poon\authormark{1}}

\address{\authormark{1} Department of Electrical and Computer Engineering, University of Toronto, 10 King's College Rd, Toronto, Ontario, M5S 3G4, Canada\\
}

\email{\authormark{*}zuyang.liu@utoronto.ca} 


\begin{abstract*} 

We benchmark Lumerical FDTD and Tidy3D for 3D simulations of passive silicon photonic components on the silicon-on-insulator~(SOI) platform.
Six devices -- including an MMI, directional coupler, waveguide crossing, mode converter, polarization splitter rotator, and ring resonator -- are simulated under matched conditions using geometries from the generic process design kit~(PDK) from GDSFactory.
Our study emphasize on comparing simulation accuracy across both solvers, alongside an analysis of runtime and broadband behavior over varying resolutions and bandwidths.
Results show that both solvers are reliable with minimal discrepancies.

\end{abstract*}

\section{Introduction}

The finite-difference time-domain~(FDTD) method is a widely used full-wave solution for modeling eletromagnetic fields in complex media and geometries~\cite{Taflove2004, Kunz1993, Teixeira2023}.
It discretizes the simulation domain into a mesh and solves Maxwell's equations at each point over time.
As a time-domain method, FDTD naturally supports broadband analysis and allows accurate modeling of arbitrarily shaped, dispersive, and subwavelength structures.
These capabilities make it especially valuable for simulating integrated photonic devices with high index contrast and complex geometries.

Due to its versatility, FDTD is employed in the simulation of photonic components including waveguides, resonators, and grating couplers~\cite{Park2008, Suzuki2007, Nguyen2013, Farmani2019, Ayata2019, Korcek2023}.
However, many devices require full three-dimensional~(3D) simulations to capture asymmetric vertical structure and polarization effects, which significantly increase computational demands.
Simulation time and memory usage scale rapidly with discretization.
As a result, a careful balance between accuracy, resolution, and runtime is essential.

Among these tools, Lumerical FDTD (Ansys) is a widely used commercial solver known for its robust performance and various built-in features with a rich desktop graphical user interface~(GUI)~\cite{ANSYS}.
Meanwhile, Tidy3D (Flexcompute) is an emerging cloud-native FDTD solver that offers graphics processing unit~(GPU) acceleration, a Python interface for streamlined workflow, and a web-based GUI~\cite{FLEXCOMPUTE}.
They differ significantly in computational performance, runtime, and user experience.
However, to our knowledge, there have not been a study that systematically compares these electromagnetic solvers under matched simulation conditions.

In this study, we benchmark Lumerical FDTD and Tidy3D for 3D FDTD simulations of a series of passive silicon photonic components on the silicon-on-insulator~(SOI) platform in terms of simulation accuracy, broadband simulation reliability, runtime, and usability.
We consider a representative set of devices including a 2$\times$2 multimode interferometer~(MMI), a directional coupler, a waveguide crossing, a mode converter, a polarization rotator splitter~(PRS), and a ring resonator.
All geometries are picked from the generic process design kit~(PDK) of GDSFactory, an open source electronics design automation~(EDA) tool for integrated circuits.~\cite{GDSFactory}.
Each device is simulated using the same material models, boundary conditions, and source configurations across both solvers.
The simulations are centered at 1550~nm.

The remainder of this manuscript is structured as follows.
Section 2 describes the simulation setup and configurations. 
Section 3 presents the simulation results and comparative analysis. 
Section 4 discusses the implications of our findings in practical photonic design workflows.
Finally, Section 5 concludes the paper.

\section{Simulation configurations}

Simulation of the same set of devices are conducted using Lumerical FDTD and Tidy3D under matched settings for comparison.
The simulation results from Lumerical FDTD are obtained from a solver version of 2025 R1 (8.33.3999), running locally using CPU on a workstation with AMD 3960X 24-core processor, 64~GB RAM, and Windows 10 OS.
Meanwhile, to demonstrate the performance on hardware more advanced than our local workstation, we also include the simulation runtime from Lumerical FDTD 2025 R2.1 that supports Ansys Cloud Burst Compute, which uses Amazon EC2 G6e instances powered by up to eight NVIDIA L40S GPUs.
Tidy3D simulations are executed on Flexcompute's GPU-accelerated cloud cluster using version 2.7.7~\cite{Minkov2024}.
The computation resources are dynamically allocated based on simulation size and hardware availability at the time.
After inquiry with Flexcompute, we were informed that one representative simulation - mode converter at the resolution of 25 cells per wavelength - was allocated to eight NVIDIA A100 GPUs.
It needs to be noted that direct wall-clock comparisons reflect the best-effort use of each platform rather than identical compute resources.
For both solvers, the geometry and simulation domain setup, boundary settings, and mesh configuration are all handled through their Python application programming interfaces~(APIs).
The Python codes are included in Code 1~\cite{Liu2025}.
The auto-shutoff threshold is set to 1$\times$10$^{-5}$ for early termination.
The simulation domain was defined to enclose the component with a distance of 2~$\mu$m from its top and bottom edges and 1~$\mu$m from edges on the device plane.
Each device is extended with 10~$\mu$m long straight waveguides at every port before importing into the solvers.
This preprocessing step is to ensure extrusion of input and output waveguides through the boundaries in the propagation direction and therefore a reliable result.
Perfectly matched layer~(PML) boundary condition is applied to all six sides of the 3D domain.
No symmetry planes are used to maintain generality.
If a diverging simulation is encountered, we change the boundary condition accordingly.
For instance, we switch to ``absorber'' in Tidy3D, which applies adiabatic absorbers containing multiple layers with gradually increasing conductivity to the boundaries.
It is suitable for dispersive materials intersecting with simulation boundaries~\cite{Flexcompute2024}.

We selected five passive silicon photonic components to span an range of mode coupling mechanisms and structure complexity.
All components are designed for a wavelength around 1550~nm and sourced directly from the generic PDK of GDSFactory.
To study the interactions of fundamental modes, we include a directional coupler and a waveguide crossing.
A 2$\times$2 MMI is simulated to evaluate multimode propagation.
Additionally, a mode converter is modeled for higher-order mode excitation.
We also select a polarization splitter rotator~(PSR) to study higher-order mode propagation and polarization handling.
Finally, we simulate a ring resonator for more complex resonance behavior.
We generate GDSII files of the components and import them to the solvers through Python APIs.

The layer stack consists of a standard SOI platform with 220~nm silicon~(Si) in top and bottom cladding of silicon oxide~(SiO$_2$).
One exception is the PSR, where the top cladding is silicon nitride~(Si$_3$N$_4$) and the bottom cladding is SiO$_2$, as demonstrated by Dai \textit{et al.}~\cite{Dai2011}
The dispersion of Si and SiO$_2$ are obtained from Handbook of Optical Constants of Solids by E. Palik \textit{et al}.~\cite{Palik1997}
The refractive index of Si$_3$N$_4$ is estimated as 2.0 around the center wavelength of 1550~nm.
On one hand, Lumerical FDTD models dispersive materials using multi-coefficient material~(MCM) model, which can describe material dispersion with higher accuracy and lower computational need~\cite{Lumerical}.
It fits the frequency-dependent permittivity of Si with two coefficients, resulting in a root mean square~(RMS) error of 2.15$\times$10$^{-5}$, without disclosing the value of fitting coefficients.
Also, the permittivity of SiO$_2$ is fitted as a constant of 2.09 in the simulation bandwidth with a RMS error of 2.43$\times$10$^{-4}$.
On the other hand, Tidy3D fits the dispersion data using the pole-residue model~\cite{Flexcompute2025}.
This model expresses the frequency-dependent permittivity as a sum of resonant material poles:
\begin{equation}
	\epsilon\left( \omega \right) = \epsilon _\infty - \sum _i \left( \frac{c_i}{j\omega+a_i} + \frac{c_i^*}{j\omega+a_i^*} \right)
\end{equation}
where $\omega$ represents the frequency, $\epsilon_\infty$ the relative permittivity at infinite frequency, and $(c_i, a_i)$ the complex pole-residue pairs.
Table~\ref{tab:pole-res} lists the fitted relative permittivity and pole-residue pairs for Si and SiO$_2$, with RMS of 1.50$\times$10$^{-2}$ and 8.63$\times$10$^{-6}$, respectively.

\begin{table}[t]
	\small
	\caption{Value of the relative permittivity and complex pole-residue pairs for modeling of Si and SiO$_2$ in Tidy3D}
	\label{tab:pole-res}
	\centering
	\begin{tabular}{|c|c|c|}
		\hline
		Material & Coefficients & Values \\
		\hline
		Si & $\epsilon_\infty$ & 10.67 \\
		& $(c_1, a_1)$ & $(-117.30-2.52\times10^{15}j, 117.28+1.33\times10^{15}j)$ \\
		& $(c_2, a_2)$ & $(-30.35-1.42\times10^{15}j, 2.38+2.91\times10^{12}j)$ \\
		\hline
		SiO$_2$ & $\epsilon_\infty$ & 1.28 \\
		& $(c_1, a_1)$ & $(-34.36-1.84\times10^{14}j, -0.014+8.87\times10^{13}j)$  \\ 
		& $(c_2, a_2)$ & $(-843.41-1.83\times10^{16}j, 713.55+7.50\times10^{15}j)$ \\ 
		\hline
	\end{tabular}
\end{table}

To evaluate the resolution-dependent performance, we applied different discretization sizes in both solvers.
The spatial resolution is determined by the number of cells per wavelength, which is the wavelength in the highest-index medium in the simulation domain.
We consider a series of different resolutions in the comparison.
No mesh override is manually applied.
In Lumerical FDTD, we choose the custom non-uniform mesh and update the minimum mesh cells per wavelength with the selected spatial resolution.
The meshing algorithm applies a smaller mesh in high index materials.
Similarly, Tidy3D provides an automatic non-uniform meshing option where the resolution is specified by the minimum steps per wavelength in each material.
Note that the automatic grid conforms to the structures in Tidy3D, which helps avoiding staircasing of permittivity at the edges.
Lumerical FDTD offers conformal mesh technology~(CMT) that can account for subcell features near the edges~\cite{AnsysCanada2025}.

Each device is excited using a mode source with bandwidth of 20~nm or 50~nm centered at 1550~nm.
The input is fundamental TE mode, except in the PRS where fundamental TM mode is considered to evaluate polarization evolution.
In Lumerical FDTD, we insert port objects at each input and output port of the devices as mode source or field and mode expansion monitor.
Accurate modeling of the source mode across the wavelength range is ensured by enabling frequency dependent profiles in the source port.
In Tidy3D, a mode source is placed at the input port and mode monitors are placed at output ports.
Additionally, field monitors are inserted in the device plane to visualize electric field distribution.
The source spectra and positions of the source and monitors are kept the same in both solvers.

\section{Simulation results and comparisons}

\subsection{Directional coupler}

\begin{figure}[htbp]
	\centering
	\includegraphics[width=0.45\linewidth]{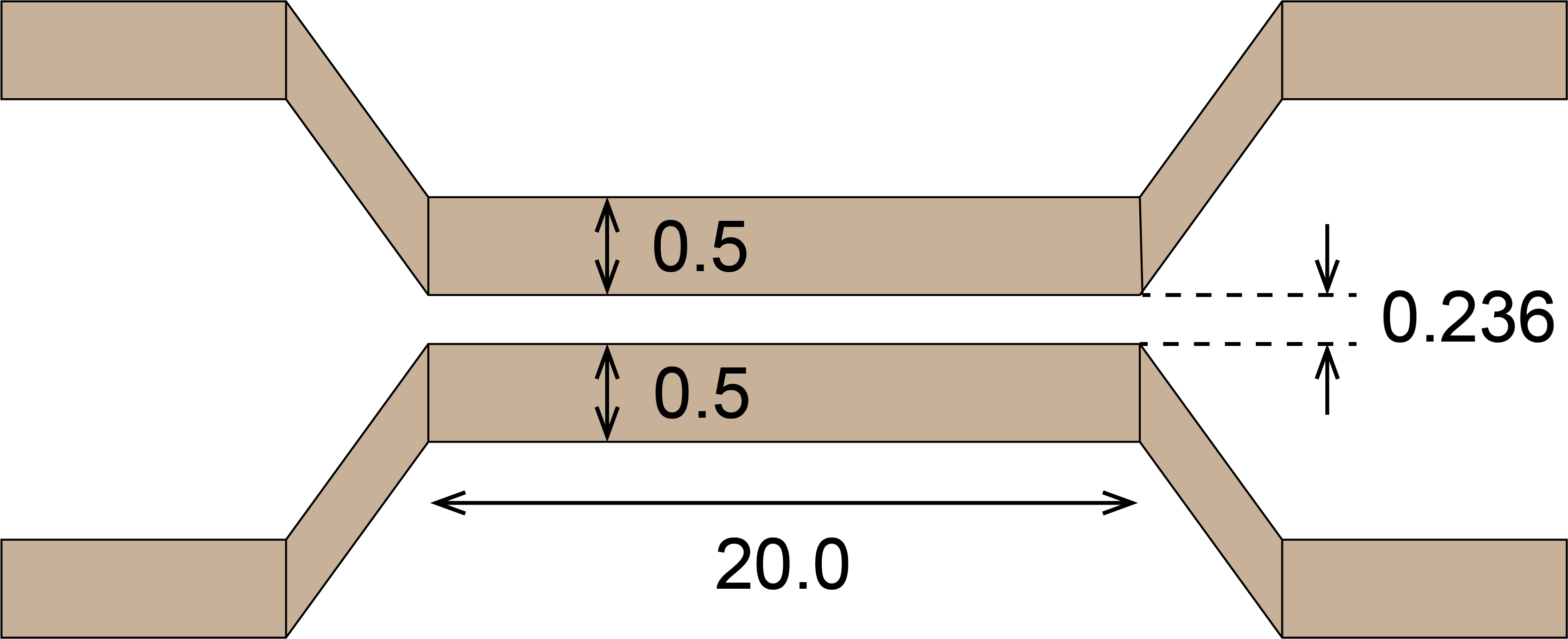}
	\caption{Sketch of the directional coupler. Dimension unit is $\mu$m.}
	\label{fig:DC_sketch}
\end{figure}

\begin{figure}[htbp]
	\centering\includegraphics[width=0.75\linewidth]{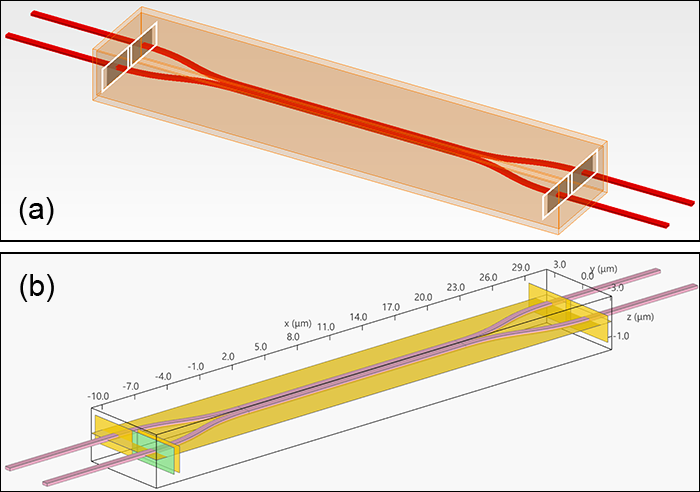}
	\caption{Schematic of the directional coupler in (a) Lumerical FDTD and (b) Tidy3D}
	\label{fig:DC_model}
\end{figure}

\begin{figure}[t]
	\centering\includegraphics[width=0.8\linewidth]{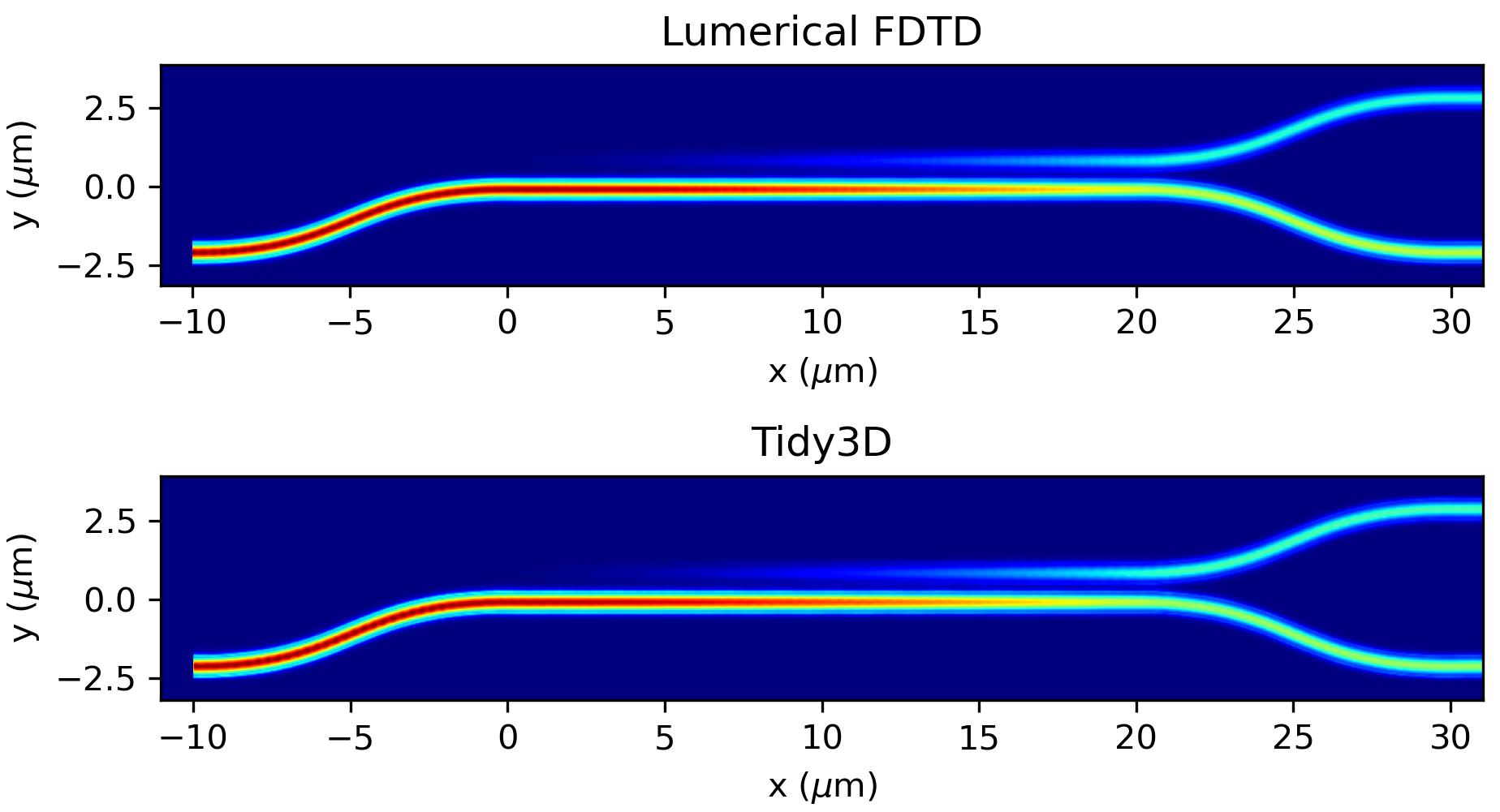}
	\caption{Electric field intensity of the directional coupler computed by Lumerical FDTD and Tidy3D at a wavelength of 1550~nm with a spatial resolution of 25 cells per wavelength.}
	\label{fig:DC_efield}
\end{figure}

The directional coupler is sourced from the \texttt{coupler} component in GDSFactory generic PDK.
As sketched in Fig.~\ref{fig:DC_sketch}, it consists of two 500~nm wide, 220~nm high Si waveguides placed in close proximity within the coupling region that is 20~$\mu$m long.
The gap width is 236~nm.
The component is cladded with SiO$_2$ on the top and the bottom.
Fig.~\ref{fig:DC_model} shows the device imported into Lumerical FDTD and Tidy3D including a 3D simulation domain, monitors on all waveguides, a 2D field monitor on the device plane, and a mode source on the lower left waveguide. 
The mode source inject fundamental TE mode at 1550~nm into the waveguide.
Fig.~\ref{fig:DC_efield} presents the normalized electric field intensity on the device plane computed by both solvers at a high resolution of 25 cells per wavelength.
As expected, a fraction of the power is coupled to the top waveguide and propagates to the cross port.
The two solvers show good agreement in the general distribution of the electric field and the intensity to each output port.

The component is simulated at a series of resolutions ranging from 6 to 25 cells per wavelength.
We notice that Tidy3D needs the boundary conditions to be set as ``absorber'' on all boundaries to avoid diverging simulation when the spatial resolution is 20 and 25.
Table~\ref{tab:DC_runtime} lists the runtime in both solvers and the peak memory usage in Lumerical FDTD, with a source bandwidth of 20~nm centered at 1550~nm.
The peak memory usage in Lumerical FDTD scales rapidly with increasing resolution due to finer discretization.
With the same size of the simulation domain~(42$\times$7$\times$4~$\mu$m$^3$), the runtime using advanced GPUs is much shorter than using our local workstation.
The two solvers show similar runtime when similar hardware is provided.
For this resolution, the total number of grid points and the elapsed simulation time and iterations are listed in Table~\ref{tab:DC_grid}. Tidy3D shows slightly finer discretization and runs for almost three times longer simulation time.

\begin{table}[htbp]
	\small
	\caption{Runtime and memory usage of directional coupler simulations with source bandwidth of 20~nm}
	\label{tab:DC_runtime}
	\centering
	\begin{tabular}{|c|c|c|c|c|}
		\hline
		Resolution & Runtime, & Local runtime, & Local memory usage, & Cloud runtime, \\
		(cells/$\lambda$) & Tidy3D~(s) & Lumerical FDTD~(s) & Lumerical FDTD~(GB) & Lumerical FDTD~(s) \\
		\hline
		6 & 18 & 148 & 0.35 & 13 \\ 
		10 & 36 & 356 & 0.92 & 19 \\ 
		15 & 63 & 1293 & 2.42 & 42 \\ 
		20 & 150 & 2606 & 5.01 & 77 \\ 
		25 & 626 & 4856 & 9.07 & 158 \\
		\hline
	\end{tabular}
\end{table}

\begin{table}[htbp]
	\small
	\caption{Total number of grid points, elapsed simulation time, and iterations of directional coupler simulations with a spatial resolution of 25 cells per wavelength.}
	\label{tab:DC_grid}
	\centering
	\begin{tabular}{|c|c|c|}
		\hline
		& Tidy3D & Lumerical FDTD \\
		\hline
		Total number of grids & 1.77$\times$10$^8$ & 9.68$\times$10$^7$\\
		Number of grids in x & 2434 & 2391 \\
		Number of grids in y & 395 & 332 \\
		Number of grids in z & 184 & 122 \\
		Elapsed simulation time (s) & 2.06$\times$10$^{-12}$ & 7.05$\times$10$^{-13}$ \\
		Elapsed iterations & 61812 & 21303 \\
		\hline
	\end{tabular}
\end{table}

\begin{figure}[htbp]
	\centering\includegraphics[width=\linewidth]{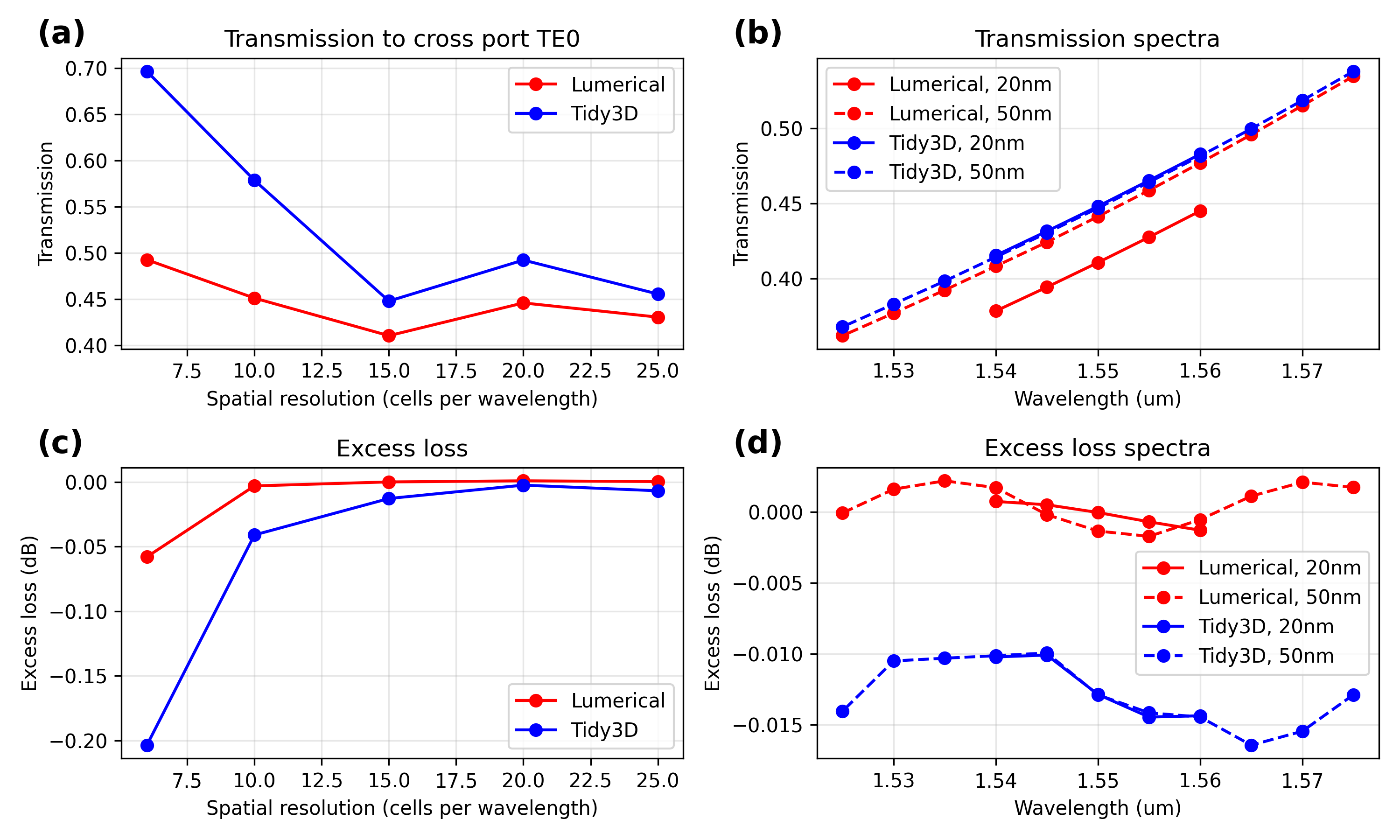}
	\caption{Simulation results of the directional coupler. (a) Transmission to the fundamental TE mode at the cross port at 1550~nm of different spatial resolutions. (b) Transmission spectra to the fundamental TE mode at the cross port with 20~nm and 50~nm source bandwidths at the spatial resolution of 15 cells per wavelength. (c) Excess loss at 1550~nm of different spatial resolutions. (d) Excess loss spectra with 20~nm and 50~nm source bandwidths at the resolution of 15 cells per wavelength.}
	\label{fig:DC_results}
\end{figure}

Then we examine the simulation result of the two solvers at different spatial resolutions.
Fig.~\ref{fig:DC_results}(a) presents the  transmission to the fundamental TE mode in the cross port at 1550~nm as a function of resolution.
These results are computed using a source bandwidth of 20~nm.
Both solvers overestimate the transmission at lower resolutions~(6 and 10 cells per wavelength) and converge to a more stable value at the resolution of 15 cells per wavelength.
Results of Tidy3D shows a more rapid decrease from 69.7$\%$ at 6 cells per wavelength to 44.8$\%$ at 15 cells per wavelength.
Meanwhile, Lumerical FDTD decreases from 49.3$\%$ at 6 cells per wavelength to 41.1$\%$ at 15 cells per wavelength.
Both solvers exhibit some fluctuations at higher resolutions beyond 15 cells per wavelength, with that in Tidy3D below 4.4$\%$ and Lumerical FDTD below 3.6$\%$.
The difference between the two solvers for the higher resolutions is up to 4.6$\%$.
These results indicate that low resolution simulations from Tidy3D is less reliable, while both solvers converge well at a moderate resolution of 15 cells per wavelength with the exact values differing slightly.

The simulations are then conducted with different source bandwidths to examine the solvers' ability of broadband simulations.
Fig.~\ref{fig:DC_results}(b) presents the transmission spectra from the two solvers with source bandwidths of 20~nm and 50~nm at a resolution of 15 cells per wavelength.
Both solvers capture the increasing transmission with wavelength.
The results from Tidy3D demonstrate good internal consistency with a deviation around 0.1$\%$ between the two bandwidths.
In contrast, Lumerical FDTD gives a noticeably lower transmission at the bandwidth of 20~nm with an offset of 3$\%$ from its 50~nm result.
The spectrum with 50~nm bandwidth is closer to the Tidy3D spectra and slightly lower than them by 0.7$\%$.
It suggests that Tidy3D is consistent across bandwidth settings, while extra attention should be taken when interpreting results with different bandwidths in Lumerical FDTD.

Fig.~\ref{fig:DC_results}(c) shows the excess loss computed for different spatial resolutions, which quantifies the intensity transmitted to the fundamental TE modes in neither of the two output ports.
Similar to the trends shown in Fig.~\ref{fig:DC_results}(a), both solvers converge to a more stable value around 15 cells per wavelength, and Tidy3D shows a sharper increase when the resolution is increased from 6 to 10 cells per wavelength.
It is demonstrated in Fig.~\ref{fig:DC_results}(d) that when the source bandwidth is increased from 20~nm to 50~nm, both solvers show good internal agreement and a very small cross-solver discrepancy of around 0.01~dB.

\subsection{Waveguide crossing}

\begin{figure}[htbp]
	\centering
	\includegraphics[width=0.6\linewidth]{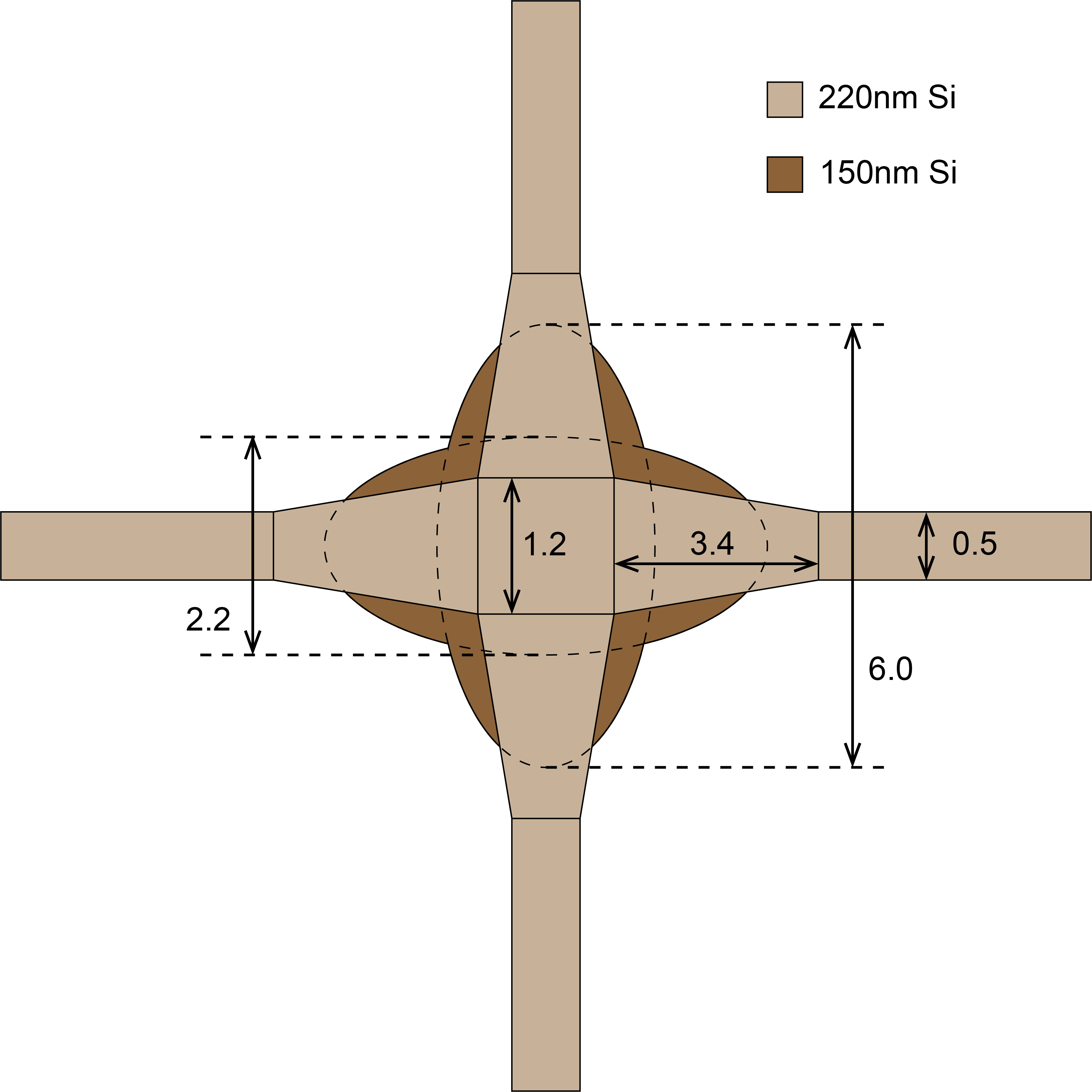}
	\caption{Sketch of the waveguide crossing. Dimension unit is $\mu$m.}
	\label{fig:CROSS_sketch}
\end{figure}

\begin{figure}[htbp]
	\centering\includegraphics[width=0.65\linewidth]{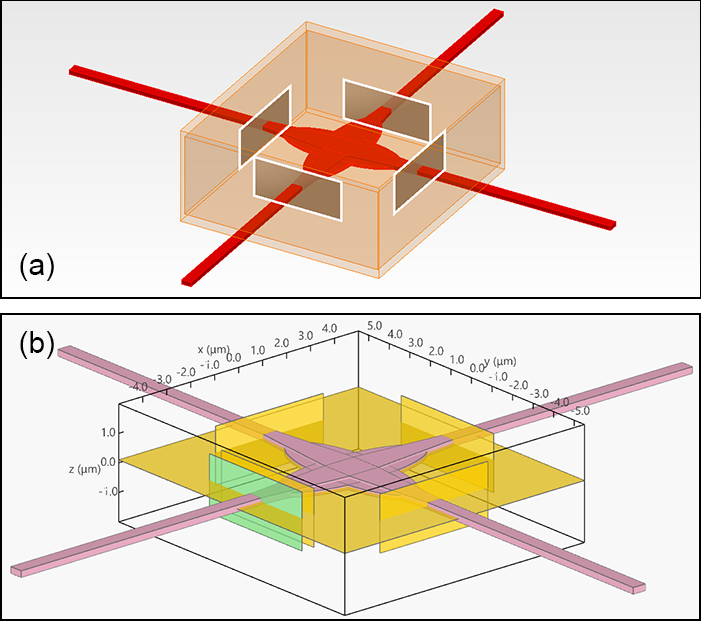}
	\caption{Schematic of the waveguide crossing in (a) Lumerical FDTD and (b) Tidy3D}
	\label{fig:CROSS_model}
\end{figure}

\begin{figure}[htbp]
	\centering\includegraphics[width=\linewidth]{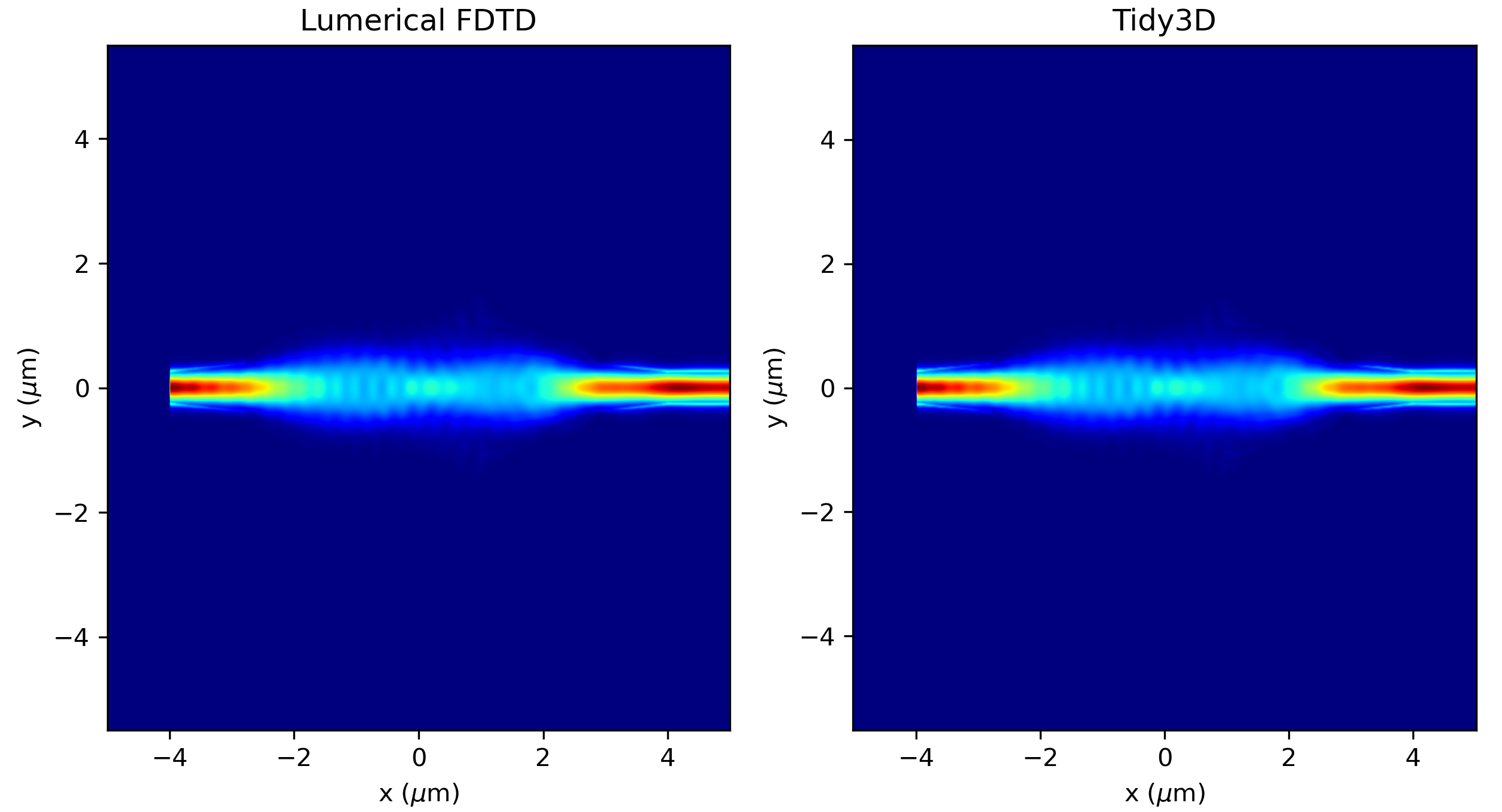}
	\caption{Electric field intensity of the waveguide crossing computed by Lumerical FDTD and Tidy3D at a wavelength of 1550~nm with a spatial resolution of 25 cells per wavelength.}
	\label{fig:CROSS_efield}
\end{figure}

The waveguide crossing is identical to the \texttt{crossing} component in GDSFactory generic PDK.
As sketched in Fig.~\ref{fig:CROSS_sketch}, it has a 1.2~$\mu$m wide Si square at the center, of which each side is joined by a 3.4~$\mu$m long taper that connects to a 500~nm wide waveguide.
The square, tapers, and waveguides are 220~nm thick.
In the slab layer that is 150~nm thick, there are two identical ellipses sharing the center of the crossing with their major axis orthogonal to each other.
The major axis is 6~$\mu$m long and the minor axis is 2.2~$\mu$m.
This component has SiO$_2$ as top and bottom cladding.
Fig.~\ref{fig:CROSS_model} demonstrates the waveguide crossing as seen in Lumerical FDTD and Tidy3D.
A fundamental TE mode is injected in the left arm at 1550~nm.
Fig.~\ref{fig:CROSS_efield} presents the electric field computed by the two solvers with a resolution of 25 cells per wavelength.
Both solvers show low-loss transmission to the through port and agree well in field patterns in the crossing section. 

We examine simulations of this waveguide crossing in a series of resolutions from 6 to 25 cells per wavelength with a source bandwidth of 20~nm.
The runtime and peak memory usage are listed in Table~\ref{tab:CROSS_runtime}.
The memory usage is lower than that for directional coupler simulations due to a smaller simulation domain.
As expected, Tidy3D is much faster than Lumerical FDTD run on our workstation.
With a simulation domain of 10$\times$11$\times$4~$\mu$m$^3$ and a resolution of 25 cells per wavelength, Tidy3D completes the simulation in 1~minute, while Lumerical completes it in 40~seconds on the cloud and 21~minutes locally.
With this resolution, the total number of grid points and the elapsed simulation time and iterations are listed in Table~\ref{tab:CROSS_grid}.
Both solvers show similar discretization, yet Tidy3D elapsed almost two times longer simulation time before the simulation reaches the auto-shutoff value.

\begin{table}[htbp]
	\small
	\caption{Runtime and memory usage of waveguide crossing simulations with source bandwidth of 20~nm}
	\label{tab:CROSS_runtime}
	\centering
	\begin{tabular}{|c|c|c|c|c|}
		\hline
		Spatial resolution & Runtime, & Local runtime, & Local memory usage, & Cloud runtime \\
		(cells/$\lambda$) & Tidy3D~(s) & Lumerical FDTD~(s) & Lumerical FDTD~(GB) & Lumerical FDTD~(s) \\
		\hline
		6 & 10 & 31 & 0.23 & 9 \\ 
		10 & 19 & 85 & 0.51 & 10 \\ 
		15 & 29 & 234 & 1.26 & 15 \\ 
		20 & 104 & 591 & 2.57 & 23 \\ 
		25 & 78 & 1268 & 4.55 & 39 \\
		\hline
	\end{tabular}
\end{table}

\begin{table}[htbp]
	\small
	\caption{Total number of grid points, elapsed simulation time, and iterations of waveguide crossing simulations with a spatial resolution of 25 cells per wavelength.}
	\label{tab:CROSS_grid}
	\centering
	\begin{tabular}{|c|c|c|}
		\hline
		& Tidy3D & Lumerical FDTD \\
		\hline
		Total number of grids & 4.83$\times$10$^7$ & 4.55$\times$10$^7$\\
		Number of grids in x & 587 & 584 \\
		Number of grids in y & 643 & 638 \\
		Number of grids in z & 128 & 122 \\
		Elapsed simulation time (s) & 6.83$\times$10$^{-13}$ & 3.59$\times$10$^{-13}$ \\
		Elapsed iterations & 20659 & 10843 \\
		\hline
	\end{tabular}
\end{table}

\begin{figure}[htbp]
	\centering\includegraphics[width=\linewidth]{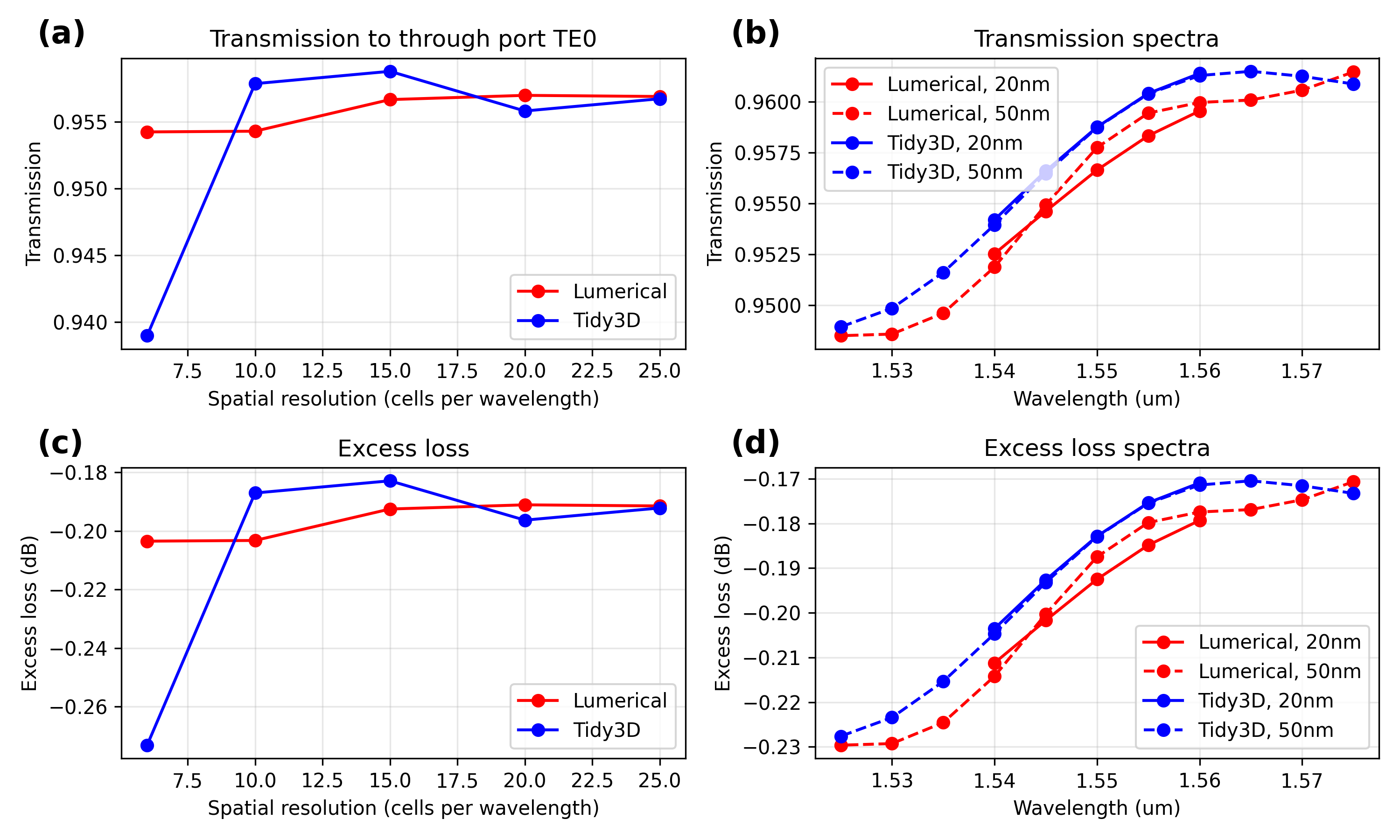}
	\caption{Simulation results of the waveguide crossing. (a) Transmission to the fundamental TE mode at the through port at 1550~nm of different spatial resolutions. (b) Transmission spectra to the fundamental TE mode at the through port with 20~nm and 50~nm source bandwidths at the spatial resolution of 15 cells per wavelength. (c) Excess loss at 1550~nm of different spatial resolutions. (d) Excess loss spectra with 20~nm and 50~nm source bandwidths at the spatial resolution of 15 cells per wavelength.}
	\label{fig:CROSS_results}
\end{figure}

We then compare the transmission to the fundamental TE mode in the through port at different spatial resolutions from 6 to 25 cells per wavelength from the two solvers, as presented in Fig.~\ref{fig:CROSS_results}(a).
It shows the transmission at 1550~nm computed with a source bandwidth of 20~nm.
The results from Lumerical FDTD is stable across the resolution range, increasing slightly from 95.4$\%$ at 6 and 10 cells per wavelength to 95.7$\%$ at resolutions not smaller than 15 cells per wavelength.
In contrast, Tidy3D increases more sharply from 93.9$\%$ at 6 cells per wavelength to a peak of 95.9$\%$ at 15 cells per wavelength, followed by a slight decline at high resolutions to meet the value of 95.7$\%$ at 25 cells per wavelength.
These results indicate that both solvers converge to within 0.3$\%$ agreement by 15 cells per wavelength, though they exhibit different convergence behaviors.
These results indicate that the two solvers converge to within 0.3$\%$ agreement by 10 cells per wavelength in Tidy3D and 15 cells per wavelength in Lumerical FDTD.
The lower resolution simulations from Tidy3D shows more deviation.
Similar trends can be seen in the comparison of excess loss in Fig.~\ref{fig:CROSS_results}(c), which quantifies the amount of power that is not coupled to the fundamental TE mode of any of the three output ports.

Fig.~\ref{fig:CROSS_results}(b) shows the transmission spectra to the fundamental TE mode in the through port with different source bandwidths in the two solvers at a resolution of 15 cells per wavelength.
Both solvers exhibit good internal consistency, with internal differences below 0.11$\%$ and 0.04$\%$ for Lumerical FDTD and Tidy3D, respectively.
The results of Lumerical FDTD is slightly lower than that of Tidy3D by an offset of 0.13$\%$.
Tidy3D demonstrates a peak of 96.15$\%$ at the wavelength of 1.565~$\mu$m, while Lumerical FDTD reaches a plateu of 96.00$\%$ around that wavelength and increases again.
Overall, the two solvers agree well across different source bandwidths at the resolution of 15 cells per wavelength.
Again, similar trends can be seen from the comparison of excess loss from the two solvers in Fig.~\ref{fig:CROSS_results}(d).

\subsection{2$\times$2 MMI}

\begin{figure}[htbp]
	\centering
	\includegraphics[width=0.75\linewidth]{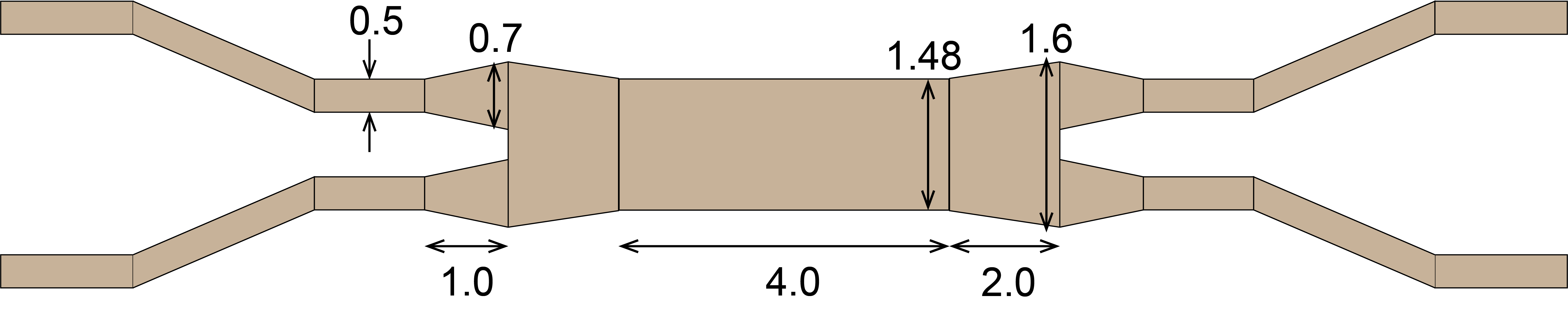}
	\caption{Sketch of the 2$\times$2 MMI. Dimension unit is $\mu$m.}
	\label{fig:MMI_sketch}
\end{figure}

\begin{figure}[htbp]
	\centering\includegraphics[width=0.65\linewidth]{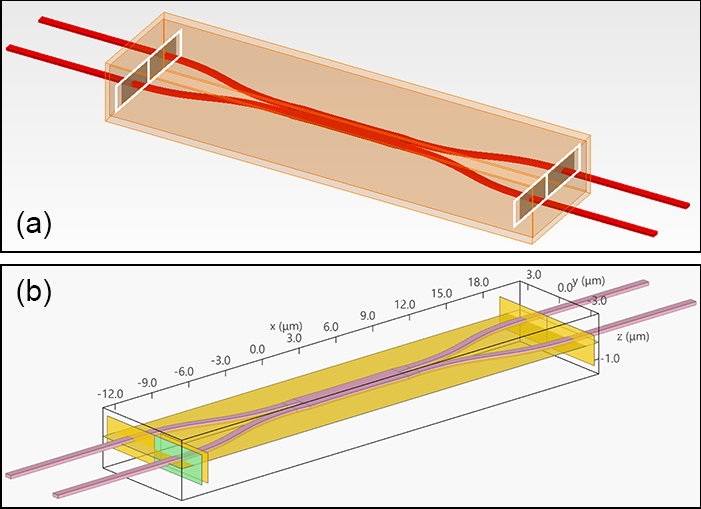}
	\caption{Schematic of the 2$\times$2 MMI in (a) Lumerical FDTD and (b) Tidy3D}
	\label{fig:MMI_model}
\end{figure}

\begin{figure}[htbp]
	\centering\includegraphics[width=\linewidth]{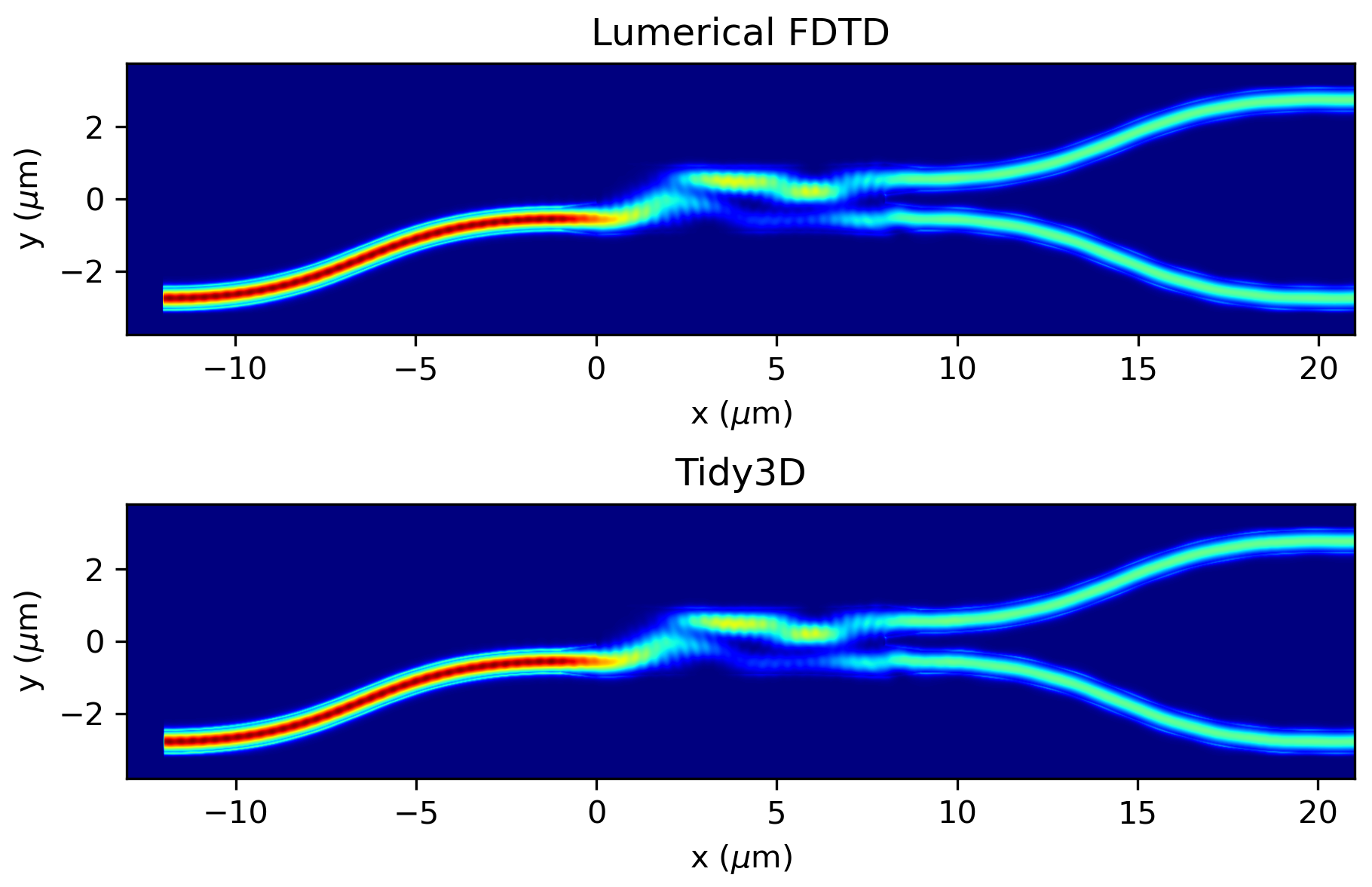}
	\caption{Electric field intensity of the 2$\times$2 MMI computed by Lumerical FDTD and Tidy3D at a wavelength of 1550~nm with a spatial resolution of 25 cells per wavelength.}
	\label{fig:MMI_efield}
\end{figure}

The MMI has the default layout provided by the \texttt{mmi2x2$\_$with$\_$sbend} component in GDSFactory generic PDK.
This layout is proposed by Guan~\textit{et al.} in \cite{Guan2017}.
The component is defined in 220~nm thick Si cladded in SiO$_2$.
The input and output waveguides are 500~nm wide and tapered to 700~nm wide before the multimode region in a length of 1~$\mu$m, as sketched in Fig.~\ref{fig:MMI_sketch}.
Fig.~\ref{fig:MMI_model} demonstrates the MMI imported to Lumerical FDTD and Tidy3D.
A fundamental TE mode at 1550~nm is injected at the lower left arm.
Examples of electric field intensity from the two solvers are presented in Fig.~\ref{fig:MMI_efield}.
They are computed at a resolution of 25 cells per wavelength.
The two solvers agree well in the multimode interference pattern and intensity distribution in the two output waveguides.

The MMI is examined in both solvers with the same set of resolutions as in the previous sections with a source bandwidth of 20~nm centered at 1550~nm.
Table~\ref{tab:MMI_runtime} lists the runtime of the both solvers and the peak memory usage in Lumerical FDTD.
The simulation domain size is 34$\times$7.5$\times$4~$\mu$m$^3$.
Table~\ref{tab:MMI_grid} lists the total number of grid points and the elapsed simulation time and iterations with the resolution of 25.
Both solvers show similar discretization while Tidy3D elapsed a 30$\%$ longer simulation time.

\begin{table}[htbp]
	\small
	\caption{Runtime and memory usage of MMI simulations with source bandwidth of 20~nm}
	\label{tab:MMI_runtime}
	\centering
	\begin{tabular}{|c|c|c|c|c|}
		\hline
		Resolution & Runtime, & Local runtime, & Local memory usage, & Cloud runtime, \\
		(cells/$\lambda$) & Tidy3D~(s) & Lumerical FDTD~(s) & Lumerical FDTD~(GB) & Lumerical FDTD~(s) \\
		\hline
		6 & 11 & 98 & 0.32 & 12\\ 
		10 & 34 & 259 & 0.82 & 15\\ 
		15 & 45 & 905 & 2.15 & 34\\ 
		20 & 63 & 2182 & 4.48 & 72\\ 
		25 & 163 & 4813 & 8.10 & 119\\
		\hline
	\end{tabular}
\end{table}

\begin{table}[htbp]
	\small
	\caption{Total number of grid points, elapsed simulation time, and iterations of MMI simulations with a spatial resolution of 25 cells per wavelength.}
	\label{tab:MMI_grid}
	\centering
	\begin{tabular}{|c|c|c|}
		\hline
		& Tidy3D & Lumerical FDTD \\
		\hline
		Total number of grids & 9.06$\times$10$^7$ & 8.59$\times$10$^7$\\
		Number of grids in x & 1929 & 1939 \\
		Number of grids in y & 367 & 363 \\
		Number of grids in z & 128 & 122 \\
		Elapsed simulation time (s) & 1.14$\times$10$^{-12}$ & 7.92$\times$10$^{-13}$ \\
		Elapsed iterations & 34262 & 23928 \\
		\hline
	\end{tabular}
\end{table}

\begin{figure}[htbp]
	\centering\includegraphics[width=\linewidth]{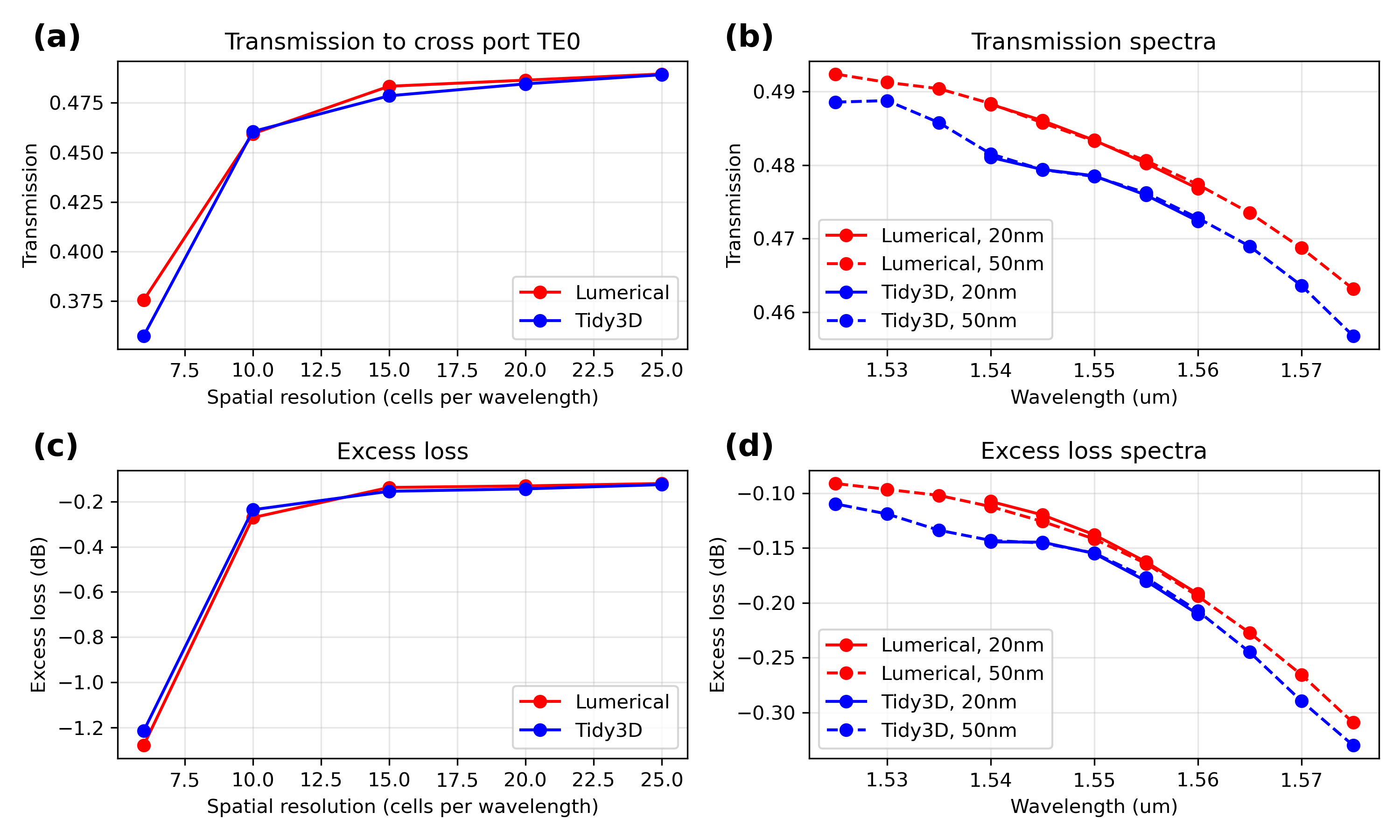}
	\caption{Simulation results of the 2$\times$2 MMI. (a) Transmission to the fundamental TE mode at the cross port at 1550~nm of different spatial resolutions. (b) Transmission spectra to the fundamental TE mode at the cross port with 20~nm and 50~nm source bandwidths at the spatial resolution of 15 cells per wavelength. (c) Excess loss at 1550~nm of different spatial resolutions. (d) Excess loss spectra with 20~nm and 50~nm source bandwidths at the spatial resolution of 15 cells per wavelength.}
	\label{fig:MMI_results}
\end{figure}

Fig.~\ref{fig:MMI_results}(a) compares the transmission to the fundamental TE mode at the cross port computed by the two solvers as a function of spatial resolution.
The results are computed using a source bandwidth of 20~nm, and only the transmission at 1550~nm is displayed.
The overall difference between the two solvers is less than 2$\%$.
In Lumerical FDTD, the transmission increases from 37.6$\%$ at 6 cells per wavelength to a more stable value around 48$\%$ at 15 cells per wavelength and beyond.
Similarly, Tidy3D increases from 35.8$\%$ at 6 cells per wavelength to 47.9$\%$ at 15 cells per wavelength.
A moderate resolution of 15 cells per wavelength is sufficient in both solvers to provide reliable results.
The good agreement and moderate convergence resolution can also be found in the quantification of excess loss as presented in Fig.~\ref{fig:MMI_results}(c).

Then we examine the performance of the two solvers with different source bandwidths.
Fig.~\ref{fig:MMI_results}(b) shows the transmission spectra to the fundamental TE mode at the cross port with bandwidths of 20~nm and 50~nm at a resolution of 15 cells per wavelength.
Both solvers exhibit excellent internal consistency across bandwidth settings and a similar decreasing transmission with increasing wavelength.
The results from Lumerical FDTD is higher than that of Tidy3D by less than 0.72$\%$.
The negligible discrepancy also applies to the excess loss presented in Fig.~\ref{fig:MMI_results}(d).

\subsection{Mode converter}

\begin{figure}[htbp]
	\centering
	\includegraphics[width=0.6\linewidth]{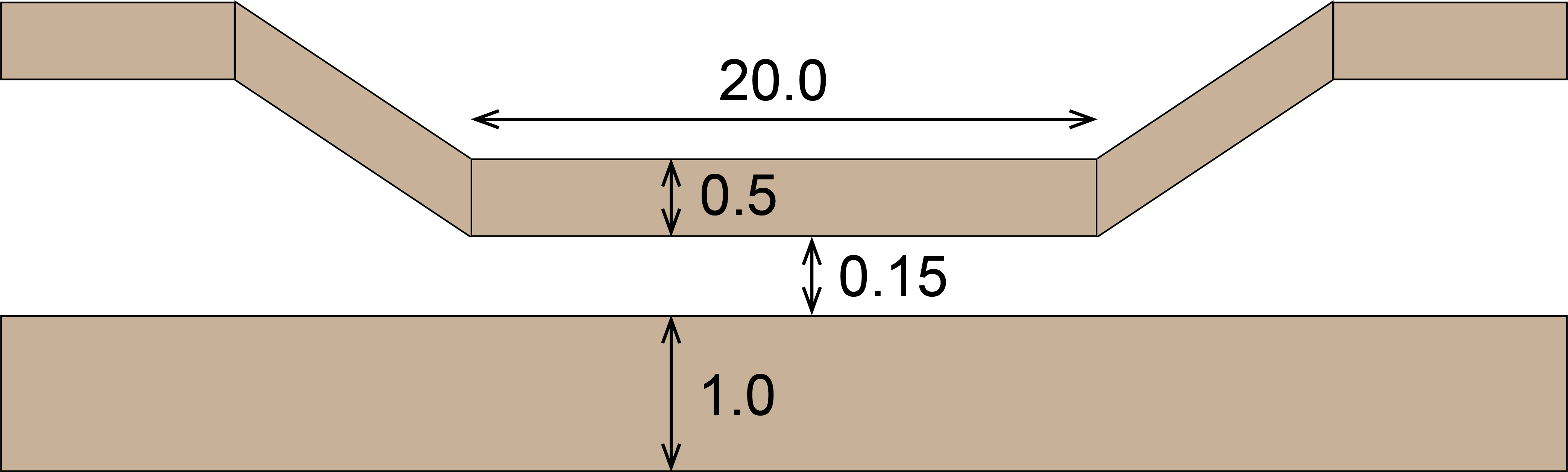}
	\caption{Sketch of the mode converter. Dimension unit is $\mu$m.}
	\label{fig:MODE_sketch}
\end{figure}

\begin{figure}[htbp]
	\centering\includegraphics[width=0.65\linewidth]{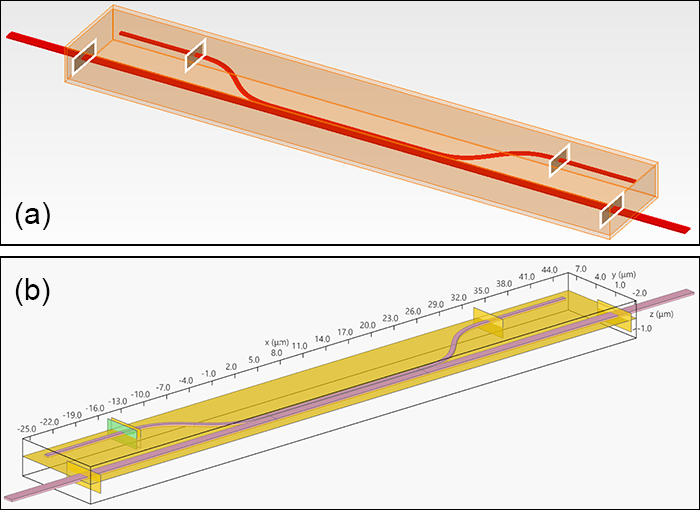}
	\caption{Schematic of the mode converter in (a) Lumerical FDTD and (b) Tidy3D}
	\label{fig:MODE_model}
\end{figure}

\begin{figure}[htbp]
	\centering\includegraphics[width=\linewidth]{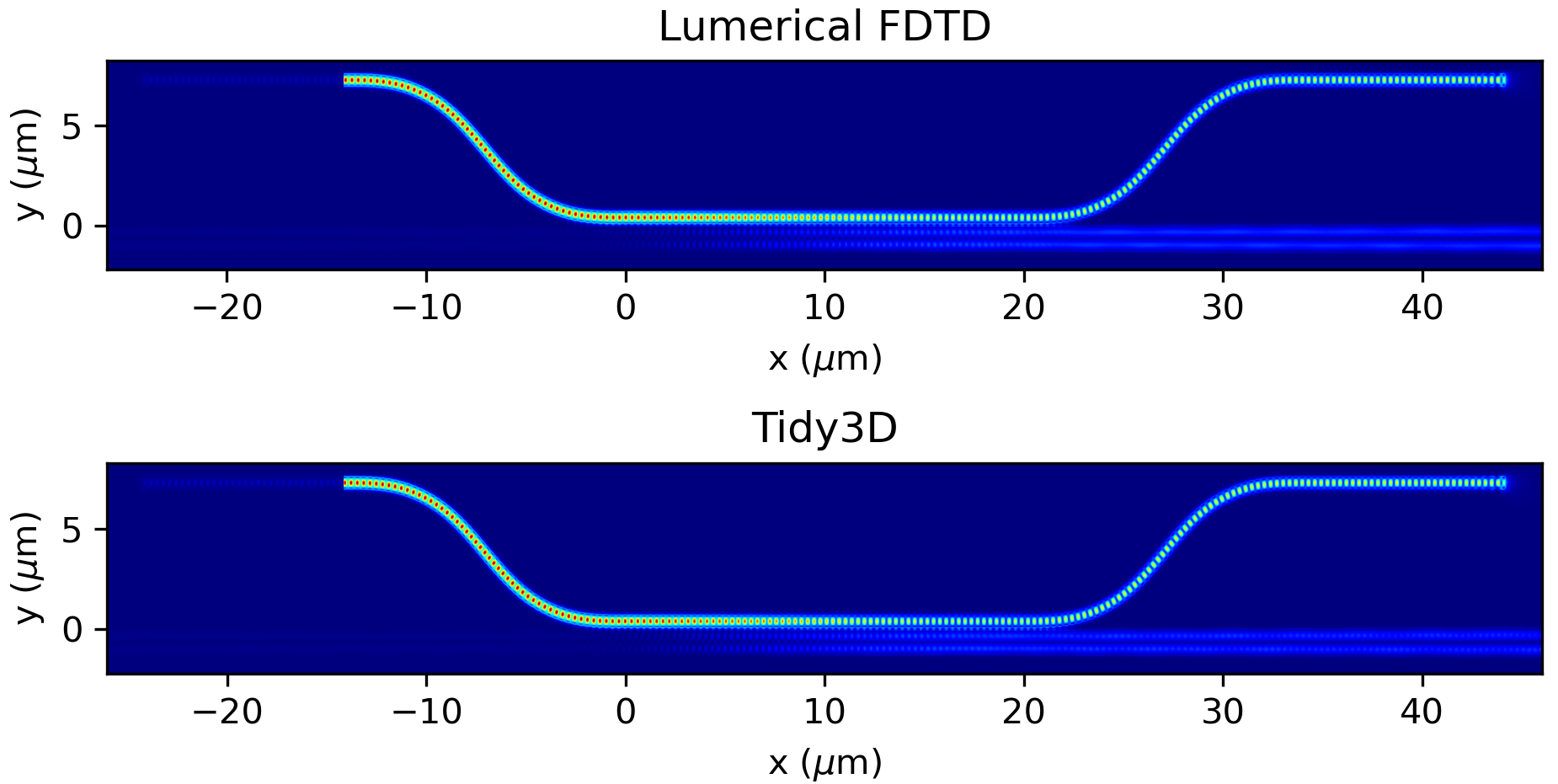}
	\caption{Electric field intensity of the mode converter computed by Lumerical FDTD and Tidy3D at a wavelength of 1550~nm with a spatial resolution of 25 cells per wavelength.}
	\label{fig:MODE_efield}
\end{figure}

The mode converter is sourced from the \texttt{mode$\_$converter} component in GDSFactory generic PDK, as proposed in~\cite{Shu2019}.
The function is to convert the fundamental TE mode in a single-mode waveguide into the first order TE mode in a wider waveguide.
It consists of a 500~nm wide waveguide and a 1~$\mu$m wide waveguide in close proximity within a 20~$\mu$m long coupling section.
The gap width is 150~nm.
The 1~$\mu$m wide waveguide is tapered to 1.2~$\mu$m wide before and after the coupling section by 25~$\mu$m long tapers.
All waveguides are made of 220~nm thick Si and cladded in SiO$_2$.
Fig.~\ref{fig:MODE_sketch} illustrates the mode converter and Fig.~\ref{fig:MODE_model} shows this component imported into Lumerical FDTD and Tidy3D.
Fig.~\ref{fig:MODE_efield} includes examples of the electric field intensity computed by Lumerical FDTD and Tidy3D at a resolution of 25 cells per wavelength.
Both solvers capture that a small fraction of the electric field is coupled from the narrow input waveguide to the wider output waveguide as a higher order mode.
It needs to be noted that Tidy3D requires the boundary condition to be ``absorber'' on all boundaries to avoid simulation diverging for the simulations of this component.

The simulation is run at a series of spatial resolutions ranging from 6 to 25 cells per wavelength with a source bandwidth of 20~nm, as in the previous sections.
The runtime in both solvers and the peak memory usage is listed in Table~\ref{tab:MODE_runtime}.
We notice that the runtime of Tidy3D does not increase monotonically with spatial resolution, as the computational speed depends on the server resource usage at the time.
The memory usage is more demanding than that in simulations of previous components due to a larger simulation domain (72$\times$10.5$\times$4~$\mu$m$^3$).
With the resolution of 25, the total number of grid points and the elapsed simulation time and iterations are listed in Table~\ref{tab:MODE_grid}.
Tidy3D shows finer discretization and elapses slightly longer simulation time.

\begin{table}[htbp]
	\small
	\caption{Runtime and memory usage of mode converter simulations with source bandwidth of 20~nm}
	\label{tab:MODE_runtime}
	\centering
	\begin{tabular}{|c|c|c|c|c|}
			\hline
			Resolution & Runtime, & Local runtime, & Local memory usage, & Cloud runtime, \\
			(cells/$\lambda$) & Tidy3D~(s) & Lumerical FDTD~(s) & Lumerical FDTD~(GB) & Lumerical FDTD~(s) \\
			\hline
			6 & 63 & 750 & 0.71 & 30 \\ 
			10 & 518 & 3599 & 2.15 & 108 \\ 
			15 & 413 & 14417 & 6.00 & 393 \\ 
			20 & 133 & 40710 & 12.90 & 384 \\ 
			25 & 2011 & 91484 & 23.63 & 810 \\
			\hline
		\end{tabular}
\end{table}

\begin{table}[htbp]
	\small
	\caption{Total number of grid points, elapsed simulation time, and iterations of mode converter simulations with a spatial resolution of 25 cells per wavelength.}
	\label{tab:MODE_grid}
	\centering
	\begin{tabular}{|c|c|c|}
			\hline
			& Tidy3D & Lumerical FDTD \\
			\hline
			Total number of grids & 4.37$\times$10$^8$ & 2.58$\times$10$^8$\\
			Number of grids in x & 4115 & 4084 \\
			Number of grids in y & 577 & 518 \\
			Number of grids in z & 184 & 122 \\
			Elapsed simulation time (s) & 5.53$\times$10$^{-12}$ & 5.18$\times$10$^{-12}$ \\
			Elapsed iterations & 165849 & 156383 \\
			\hline
		\end{tabular}
\end{table}

\begin{figure}[htbp]
	\centering\includegraphics[width=\linewidth]{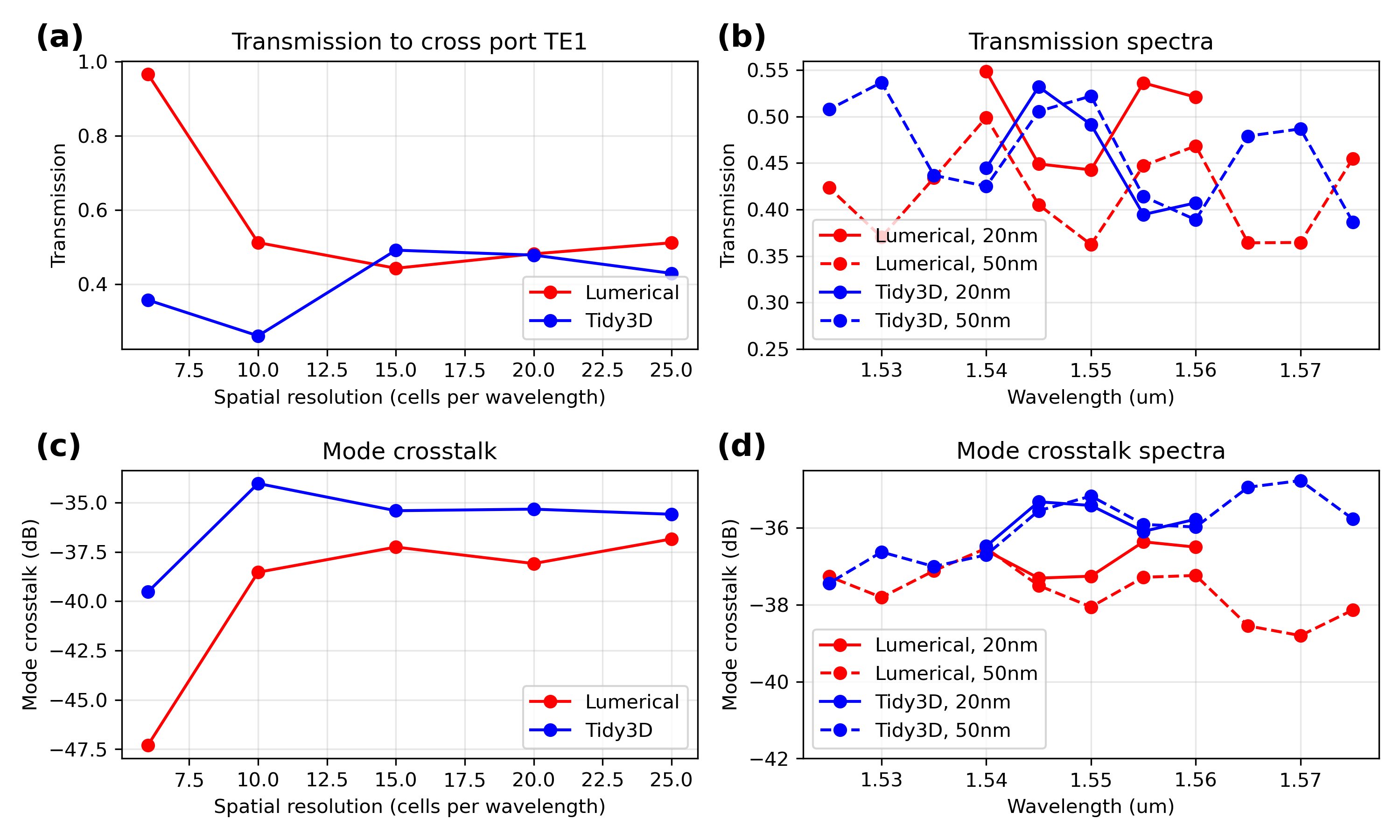}
	\caption{Simulation results of the mode converter. (a) Transmission to the TE1 mode at the cross port at 1550~nm of different spatial resolutions. (b) Transmission spectra to the TE1 mode at the cross port with 20~nm and 50~nm source bandwidths at the spatial resolution of 15 cells per wavelength. (c) Mode crosstalk at 1550~nm of different spatial resolutions. (d) Mode crosstalk spectra with 20~nm and 50~nm source bandwidths at the spatial resolution of 15 cells per wavelength.}
	\label{fig:MODE_results}
\end{figure}

We compare the transmission to the first order TE mode at the cross port across the spatial resolutions from the two solvers, as presented in Fig.~\ref{fig:MODE_results}(a).
It considers the transmission at 1550~nm computed with a source bandwidth of 20~nm.
In Lumerical FDTD, the transmission drops sharply from an unphysical value of 96.7$\%$ at 6 cells per wavelength to more realistic values around 50$\%$ for higher resolutions above 10 cells per wavelength.
In contrast, the transmission from Tidy3D increases gradually from 35.7$\%$ at 6 cells per wavelength to 49.2$\%$ at 15 cells per wavelength.
The two solvers agree well between 15 and 20 cells per wavelength, with discrepancies below 5$\%$.
These results suggest that Lumerical FDTD converges to approximately 50$\%$ at a slightly lower resolution of 10 cells per wavelength, while Tidy3D converges to a similar value at 15 cells per wavelength.
Similar trends can be seen from Fig.~\ref{fig:MODE_results}(c) showing the mode crosstalk, which quantifies the transmission from the source mode to the fundamental TE mode in the cross port.

The bandwidth-dependent performance is evaluated by comparing transmission spectra to the first order TE mode at the cross port with 20~nm and 50~nm source bandwidths in both solvers at a resolution of 15 cells per wavelength.
The results are presented in Fig.~\ref{fig:MODE_results}(b).
Both solvers give substantial variations across the spectral range.
Tidy3D shows internal consistency with variation below 3$\%$ between the 20~nm and 50~nm spectra.
In Lumerical  FDTD, the internal deviation is up to 9$\%$.
The strong variations across the spectral range can also be seen in the mode crosstalk spectra for both solvers presented in Fig.~\ref{fig:MODE_results}(d).
Lumerical FDTD shows an internal deviation of around 1~dB in the mode crosstalk, indicating that special attention should be taken when interpreting results from different bandwidth settings.

\begin{figure}[htbp]
	\centering\includegraphics[width=\linewidth]{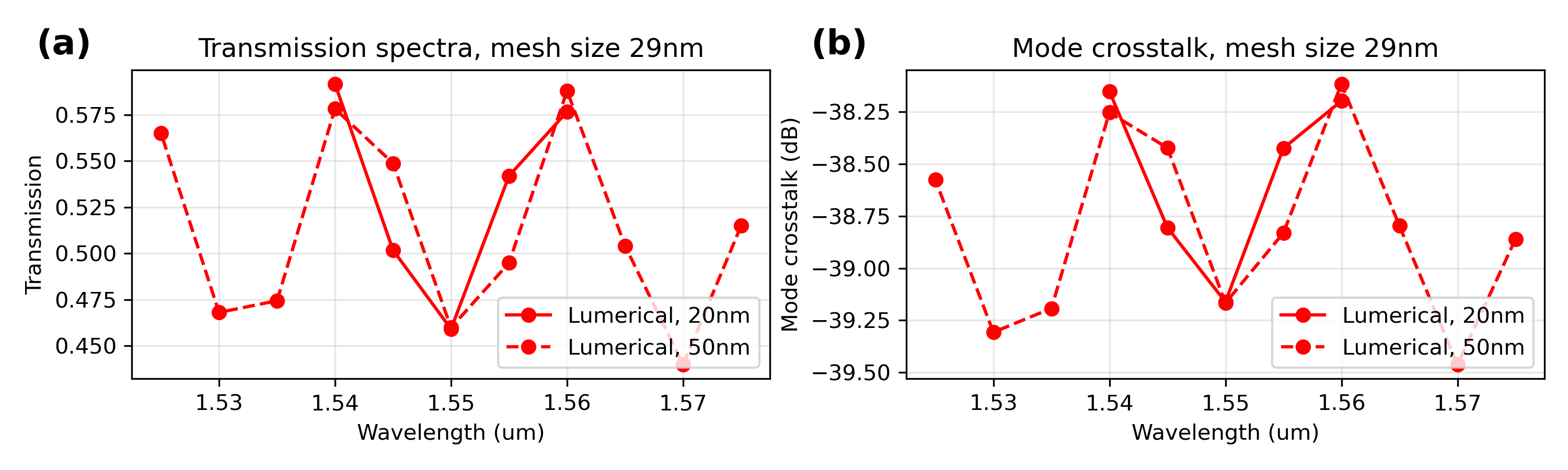}
	\caption{Simulation results of the mode converter in Lumerical FDTD with a mesh size of 29~nm. (a) Transmission spectra to the TE1 mode in the cross port. (b) Mode crosstalk spectra in the cross port.}
	\label{fig:MODE_mesh}
\end{figure}

When using the custom non-uniform mesh in Lumerical FDTD, the mesh size is determined by the lower bound of the wavelength range, the highest refractive index in the simulation domain, and the specified number of cells per wavelength.
It causes a small difference in the mesh size: 29.4~nm for 20~nm source bandwidth and 29.2~nm for 50~nm source bandwidth.
By changing the mesh size to a constant value of 29~nm, the discrepancy between simulations with different bandwidths can be significantly reduced to below 5$\%$, as shown in Fig.~\ref{fig:MODE_mesh}(a) and (b).
Comparing to the transmission spectra from Tidy3D in Fig.~\ref{fig:MODE_results}(b), there is still a large discrepancy over 16$\%$ due to mismatched positions of peaks and valleys.

\subsection{Polarization splitter rotator}

\begin{figure}[htbp]
	\centering
	\includegraphics[width=0.75\linewidth]{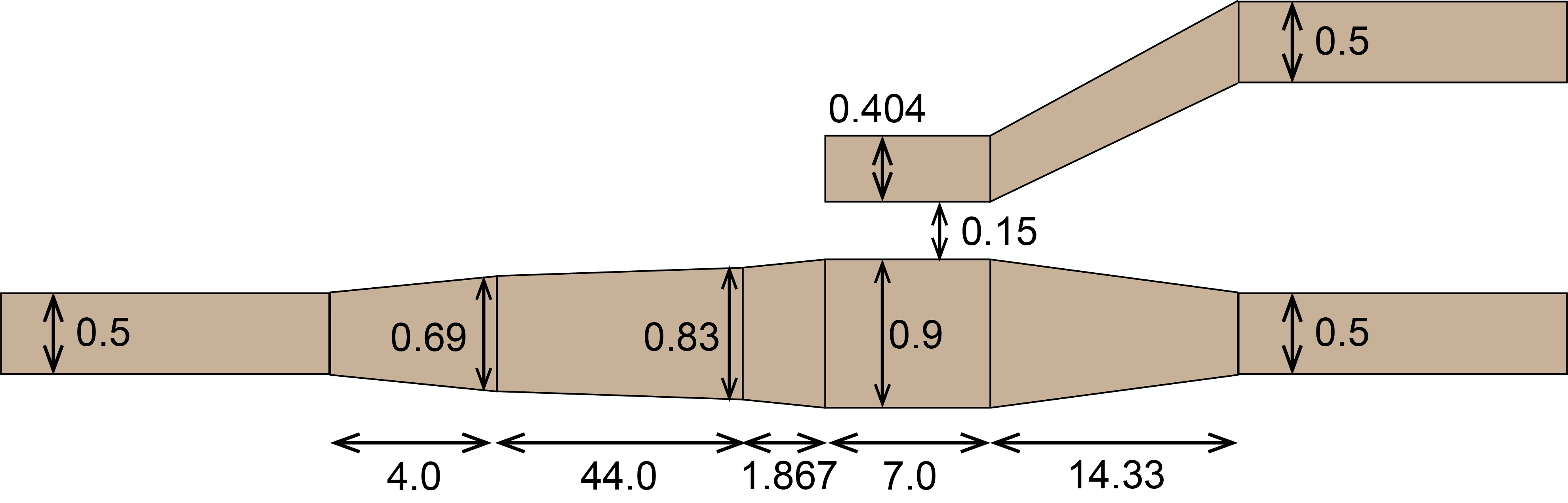}
	\caption{Sketch of the PSR. Dimension unit is $\mu$m.}
	\label{fig:PSR_sketch}
\end{figure}

\begin{figure}[htbp]
	\centering\includegraphics[width=0.65\linewidth]{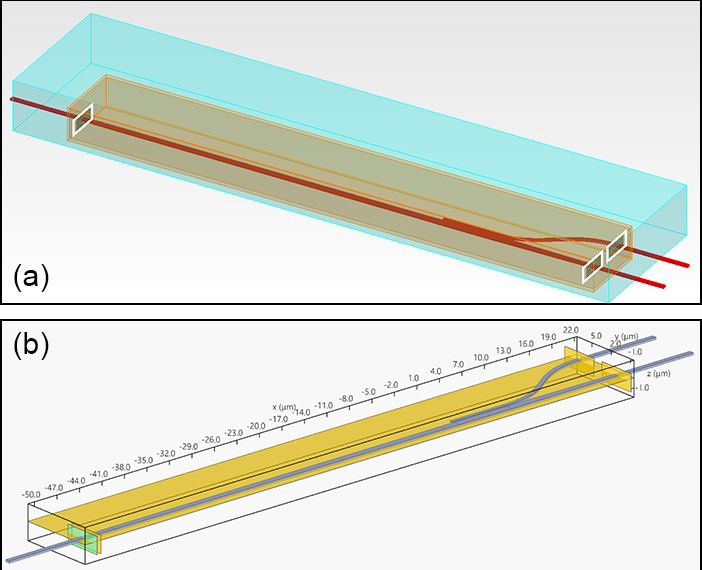}
	\caption{Schematic of the PSR in (a) Lumerical FDTD and (b) Tidy3D}
	\label{fig:PSR_model}
\end{figure}

\begin{figure}[htbp]
	\centering\includegraphics[width=\linewidth]{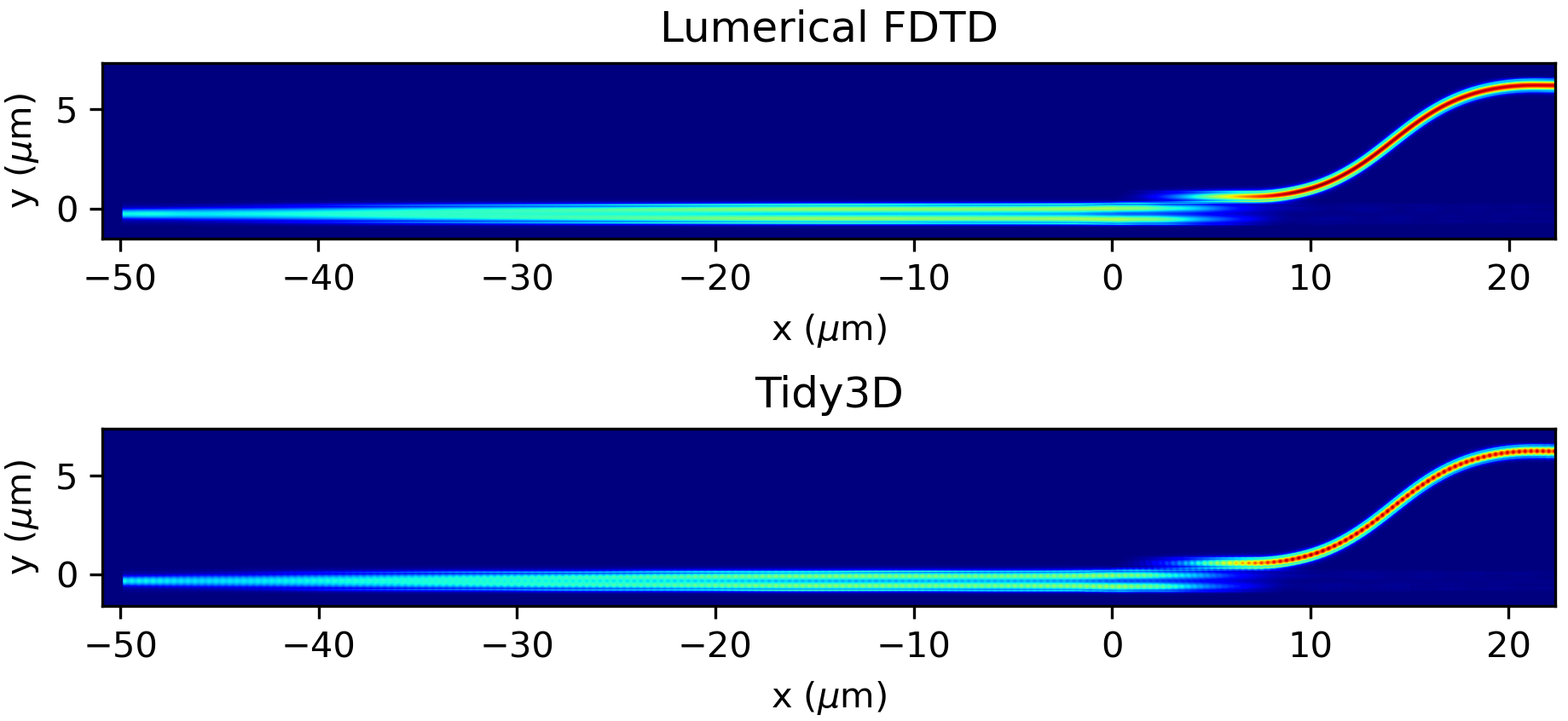}
	\caption{Electric field intensity of the PSR computed by Lumerical FDTD and Tidy3D at a wavelength of 1550~nm with a spatial resolution of 25 cells per wavelength.}
	\label{fig:PSR_efield}
\end{figure}

The PRS is identical to the \texttt{polarization$\_$splitter$\_$rotator} component in GDSFactory generic PDK.
The device converts the injected fundamental TM mode into higher order TE mode and efficiently couple it to the fundamental TE mode of another waveguide.
It can be used for polarization (de-)multiplexing in telecommunications.
In this design, the component is defined in 220~nm thick Si, with SiO$_2$ as bottom cladding and Si$_3$N$_4$ as top cladding.
The refractive index of Si$_3$N$_4$ is approximated to 2.0 in the simulation bandwidth, as presented in~\cite{Dai2011}.
Fig.~\ref{fig:PSR_sketch} sketches the PSR with its key dimensions and Fig.~\ref{fig:PSR_model} shows the component imported into Lumerical FDTD and Tidy3D.
The input port on the left and the two output ports on the right are extended with 10~$\mu$m long, 500~nm wide waveguides before importing into Lumerical FDTD and Tidy3D.
A mode source is added on the left that inject fundamental TM mode at 1550~nm into the PSR.
Fig.~\ref{fig:PSR_efield} presents the electric field intensity computed by the two solvers at a high resolution of 25 cells per wavelength, both showing efficient coupling from the left port to the upper right port.
Tidy3D shows some standing wave patterns over the entire component.

The PSR is simulated at a range of resolutions from 6 to 25 cells per wavelength.
Table~\ref{tab:PSR_runtime} lists the runtime of the two solvers and the peak memory usage of Lumerical FDTD.
Again, Tidy3D completes the simulations faster than Lumerical FDTD.
In this case, the simulation domain size is 73.2$\times$8.9$\times$4~$\mu$m$^3$.
The total number of grid points and the elapsed simulation time and iterations are listed in Table~\ref{tab:PSR_grid}.
Tidy3D shows finer discretization and elapses almost two times longer simulation time.

\begin{table}[htbp]
	\small
	\caption{Runtime and memory usage of PSR simulations with source bandwidth of 20~nm}
	\label{tab:PSR_runtime}
	\centering
	\begin{tabular}{|c|c|c|c|c|}
			\hline
			Resolution & Runtime, & Local runtime, & Local memory usage, & Cloud runtime,\\
			(cells/$\lambda$) & Tidy3D~(s) & Lumerical FDTD~(s) & Lumerical FDTD~(GB) & Lumerical FDTD~(s) \\
			\hline
			6 & 22 & 303 & 0.66 & 25 \\ 
			10 & 65 & 1073 & 2.14 & 58 \\ 
			15 & 81 & 3478 & 6.05 & 150 \\ 
			20 & 613 & 9665 & 13.06 & 277 \\ 
			25 & 1271 & 21896 & 24.01 & 481 \\
			\hline
		\end{tabular}
\end{table}

\begin{table}[htbp]
	\small
	\caption{Total number of grid points, elapsed simulation time, and iterations of PSR simulations with a spatial resolution of 25 cells per wavelength.}
	\label{tab:PSR_grid}
	\centering
	\begin{tabular}{|c|c|c|}
			\hline
			& Tidy3D & Lumerical FDTD \\
			\hline
			Total number of grids & 4.38$\times$10$^8$ & 2.63$\times$10$^8$\\
			Number of grids in x & 4182 & 4152 \\
			Number of grids in y & 524 & 463 \\
			Number of grids in z & 200 & 137 \\
			Elapsed simulation time (s) & 2.50$\times$10$^{-12}$ & 1.22$\times$10$^{-12}$ \\
			Elapsed iterations & 74879 & 36980 \\
			\hline
		\end{tabular}
\end{table}

\begin{figure}[htbp]
	\centering\includegraphics[width=\linewidth]{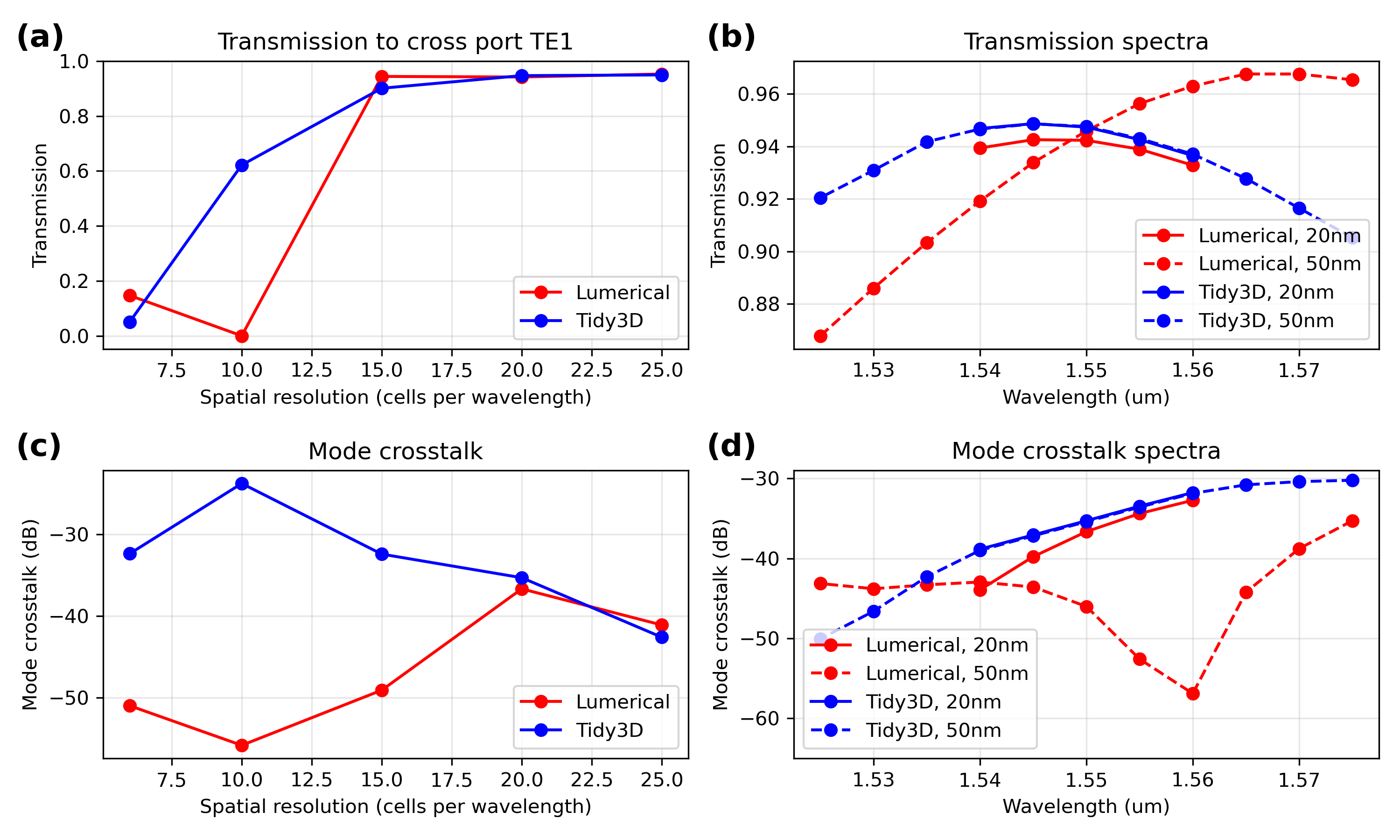}
	\caption{Simulation results of the PSR. (a) Transmission to the fundamental TE mode at the upper port at 1550~nm of different spatial resolutions. (b) Transmission spectra to the fundamental TE mode at the upper port with 20~nm and 50~nm source bandwidths at the spatial resolution of 20 cells per wavelength. (c) Mode crosstalk at 1550~nm of different spatial resolutions. (d) Mode crosstalk spectra with 20~nm and 50~nm source bandwidths at the spatial resolution of 20 cells per wavelength.}
	\label{fig:PSR_results}
\end{figure}

Fig.~\ref{fig:PSR_results}(a) plot the transmission to the fundamental TE mode at the upper right port at 1550~nm as a function of resolution, simulated with a source bandwidth of 20~nm.
Tidy3D shows a gradual increase from 5.1$\%$ at 6 cells per wavelength to 90.1$\%$ at 15 cells per wavelength.
In contrast, Lumerical FDTD starts from 14.7$\%$ at 6 cells per wavelength, followed by a dip to 0.0$\%$, and then sharply increases to 94$\%$ at 15 cells per wavelength.
Both solvers converge to a stable transmission around 95$\%$ but at different resolutions: Lumerical FDTD at 15 cells per wavelength and Tidy3D at 20 cells per wavelength.
The difference between their stable value is below 0.5$\%$.
Fig.~\ref{fig:PSR_results}(c) shows the mode crosstalk (transmission to the fundamental TM mode at the upper right port) at 1550~nm.
The two solvers only agree above 20 cells per wavelength.
These results suggest that the two solvers agree well when the resolution is sufficiently high, and Lumerical FDTD converges slightly faster in the simulation of the PSR.

Next, the simulation is examined with different source bandwidths at a resolution of 20 cells per wavelength, where the transmission in both solvers converges.
Fig.~\ref{fig:PSR_results}(b) presents the transmission spectra computed by Tidy3D and Lumerical FDTD.
Tidy3D shows excellent internal consistency with less than 0.1$\%$ variation between the results with 20~nm and 50~nm bandwidths.
However, Lumerical FDTD gives significantly different spectra between these two bandwidth settings.
The spectrum with 20~nm bandwidth peaks around 1545~nm with a maximum of 93.9$\%$, similar to Tidy3D results with an offset below 0.7$\%$.
Meanwhile, the one with 50~nm bandwidth reaches a maximum of 96.8$\%$ at 1565~nm.
The peak wavelength is shifted by 20~nm.
In the mode crosstalk spectra presented in Fig.~\ref{fig:PSR_results}(d), there is also a considerable internal deviation for Lumerical FDTD.
A minimum of -56.9~dB can be found around 1.56~$\mu$m in the spectrum with 50~nm bandwidth.
As the material dispersion and frequency dependent mode behavior are properly modeled in both solvers, it suggests an underlying problem with broadband simulations in Lumerical FDTD.

\begin{figure}[htbp]
	\centering\includegraphics[width=\linewidth]{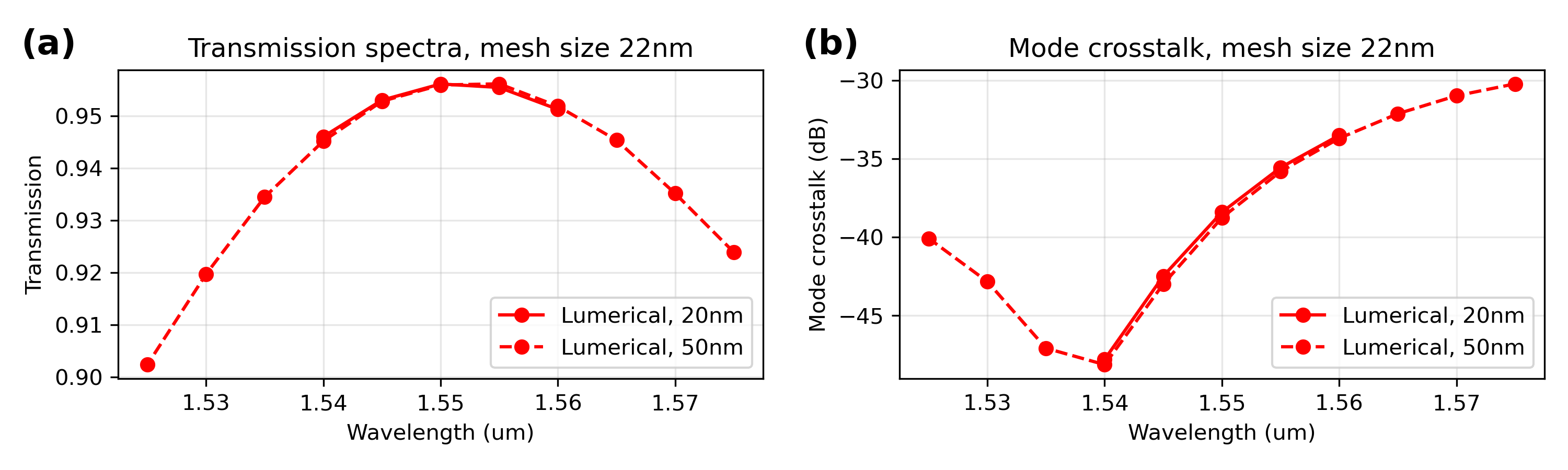}
	\caption{Simulation results of the PSR in Lumerical FDTD with a mesh size of 22~nm. (a) Transmission spectra to the fundamental TE mode in the upper port. (b) Mode crosstalk spectra in the upper port.}
	\label{fig:PSR_mesh}
\end{figure}

Similar to the analysis of the mode converter, there is a small difference of mesh size caused by varied source bandwidths in Lumerical FDTD.
Given the resolution of 20~cells per wavelength, the mesh size is 22.1~nm and 21.9~nm for 20~nm and 50~nm bandwidths, respectively.
By changing the mesh size to a constant value of 22~nm, the discrepancy between simulations with different bandwidths is almost eliminated, as shown in Fig.~\ref{fig:PSR_mesh}(a) and (b).
The difference in transmission is reduced to 0.06$\%$.
For the mode crosstalk in Fig.~\ref{fig:PSR_mesh}(b), the difference is 0.5~dB and the minimum moves to 1.54~$\mu$m and increases to -48.1~dB.
Comparing the transmission spectra in Fig.~\ref{fig:PSR_mesh}(a) to spectra from Tidy3D in Fig.~\ref{fig:PSR_results}(b), the inter-solver discrepancy is reduced to lower than 2$\%$ over the 50~nm bandwidth.

\subsection{Ring resonator}

\begin{figure}[htbp]
	\centering
	\includegraphics[width=0.6\linewidth]{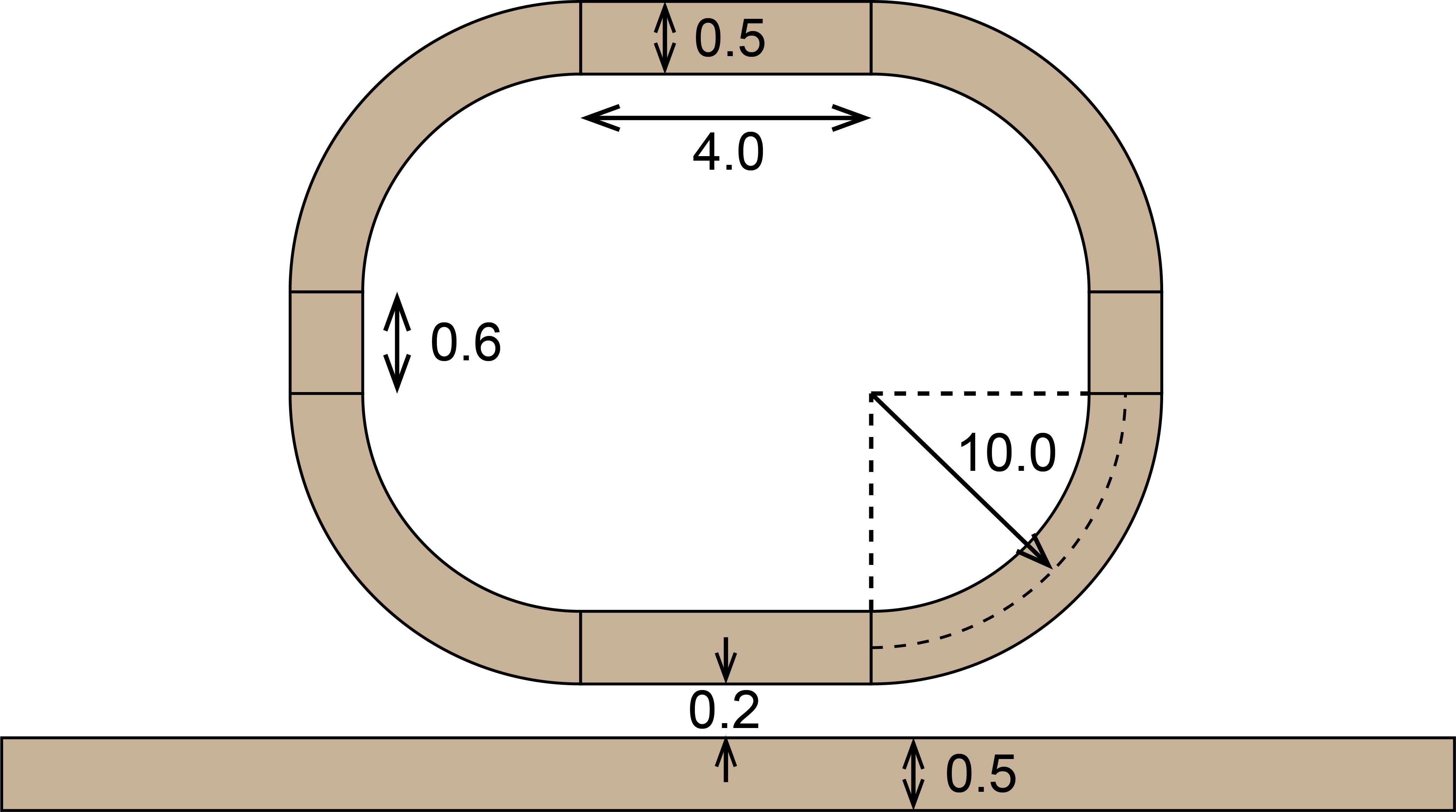}
	\caption{Sketch of the ring resonator. Dimension unit is $\mu$m.}
	\label{fig:RING_sketch}
\end{figure}

\begin{figure}[htbp]
	\centering\includegraphics[width=0.65\linewidth]{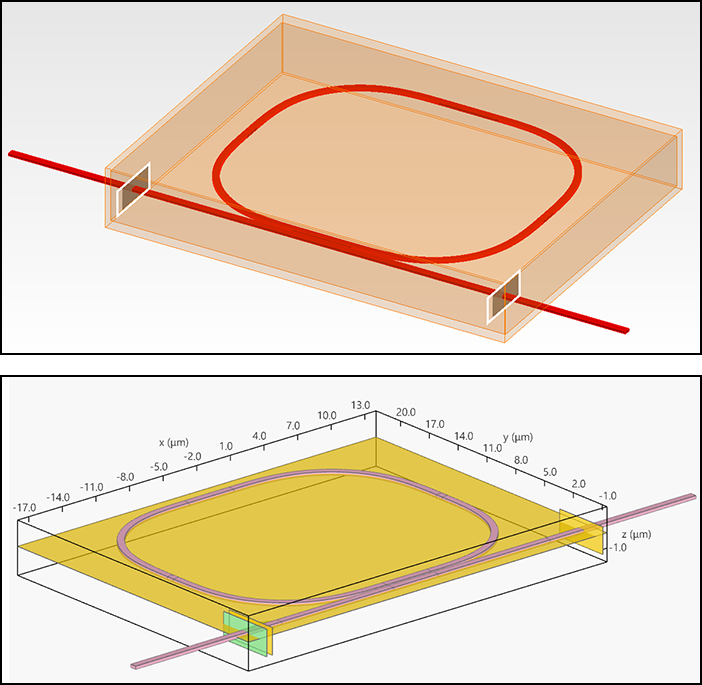}
	\caption{Schematic of the ring resonator in (a) Lumerical FDTD and (b) Tidy3D}
	\label{fig:RING_model}
\end{figure}

\begin{figure}[htbp]
	\centering\includegraphics[width=\linewidth]{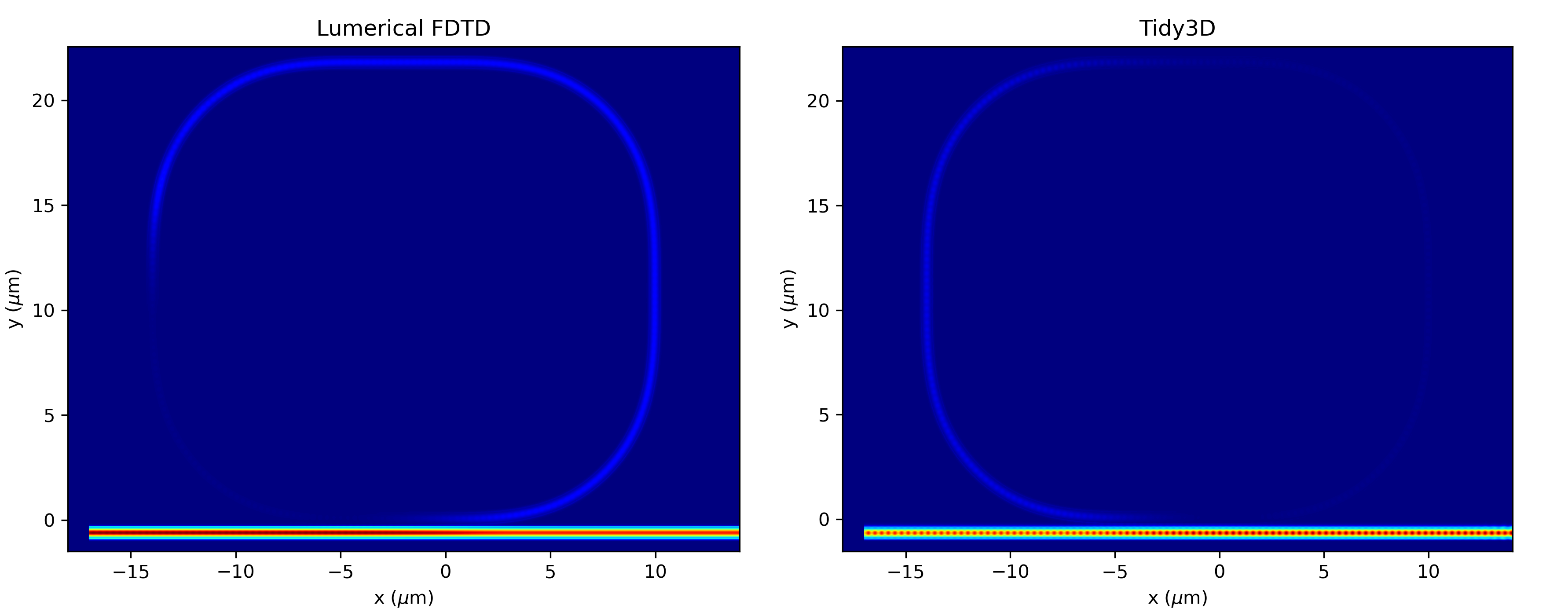}
	\caption{Electric field intensity of the ring resonator computed by Lumerical FDTD and Tidy3D at a wavelength of 1550~nm with a spatial resolution of 25 cells per wavelength.}
	\label{fig:RING_efield}
\end{figure}

The ring resonator uses the default layout of \texttt{ring$\_$single} in GDSFactory generic PDK.
This component can serve as an optical filter.
As sketched in Fig.~\ref{fig:RING_sketch}, it consists of a 4.0~$\mu$m-long directional coupler at the bottom with a gap width of 0.2~$\mu$m, two 0.6~$\mu$m-long vertical straight waveguides on the left and right, a 4.0~$\mu$m-long horizontal straight waveguide at the top, and four bends connecting them with a radius of 10~$\mu$m.
The waveguide width is 500~nm.
Fig.~\ref{fig:RING_model} shows the component imported into Lumerical FDTD and Tidy3D, with the input and output ports extended with 10~$\mu$m long, 500~nm wide waveguides.
A mode source is added on the left that inject the fundamental TE mode into the ring resonator with source bandwidths of 20~nm and 50~nm centered at 1550~nm.
Fig.~\ref{fig:RING_efield} shows the electric field intensity computed by the two solvers at 1550~nm for a resolution of 25 cells per wavelength, where the intensity distribution differs a little between the two solvers.

We simulate the ring resonator for spatial resolutions varied from 6 to 25 cells per wavelength.
The simulation domain size for the ring resonator is 32$\times$24.05$\times$4~$\mu$m$^3$.
The runtime in the two solvers and the peak memory usage in Lumerical FDTD are listed in Table~\ref{tab:RING_runtime}.
As for previous components, Tidy3D is faster than Lumerical FDTD.
Table~\ref{tab:RING_grid} lists the number of grids and the elapsed simulation time and iterations at the highest resolution of 25 cells per wavelength.
Tidy3D applies slightly finer discretization along all three dimensions.

\begin{table}[htbp]
	\small
	\caption{Runtime and memory usage of ring resonator simulations with source bandwidth of 20~nm}
	\label{tab:RING_runtime}
	\centering
	\begin{tabular}{|c|c|c|c|c|}
		\hline
		Resolution & Runtime, & Local runtime, & Local memory usage, & Cloud runtime, \\
		(cells/$\lambda$) & Tidy3D~(s) & Lumerical FDTD~(s) & Lumerical FDTD~(GB) & Lumerical FDTD~(s) \\
		\hline
		6  & 233 & 1168 & 3.14 & 52 \\ 
		10 & 307 & 4685 & 8.36 & 157 \\ 
		15 & 1104 & 16815 & 18.50 & 476 \\ 
		20 & 2594 & 44530 & 32.62 & 732 \\ 
		25 & 2819 & 101191 & 49.42 & 1433 \\
		\hline
	\end{tabular}
\end{table}

\begin{table}[htbp]
	\small
	\caption{Total number of grid points, elapsed simulation time, and iterations of ring resonator simulations with a spatial resolution of 25 cells per wavelength.}
	\label{tab:RING_grid}
	\centering
	\begin{tabular}{|c|c|c|}
		\hline
		& Tidy3D & Lumerical FDTD \\
		\hline
		Total number of grids & 4.68$\times$10$^8$ & 2.90$\times$10$^8$ \\
		Number of grids in x & 1873 & 1826  \\
		Number of grids in y & 1358 & 1302  \\
		Number of grids in z & 184 & 122 \\
		Elapsed simulation time (s) & 6.40$\times$10$^{-12}$  & 6.40$\times$10$^{-12}$ \\
		Elapsed iterations & 1.92$\times$10$^5$ & 1.95$\times$10$^5$ \\
		\hline
	\end{tabular}
\end{table}

\begin{figure}[htbp]
	\centering\includegraphics[width=\linewidth]{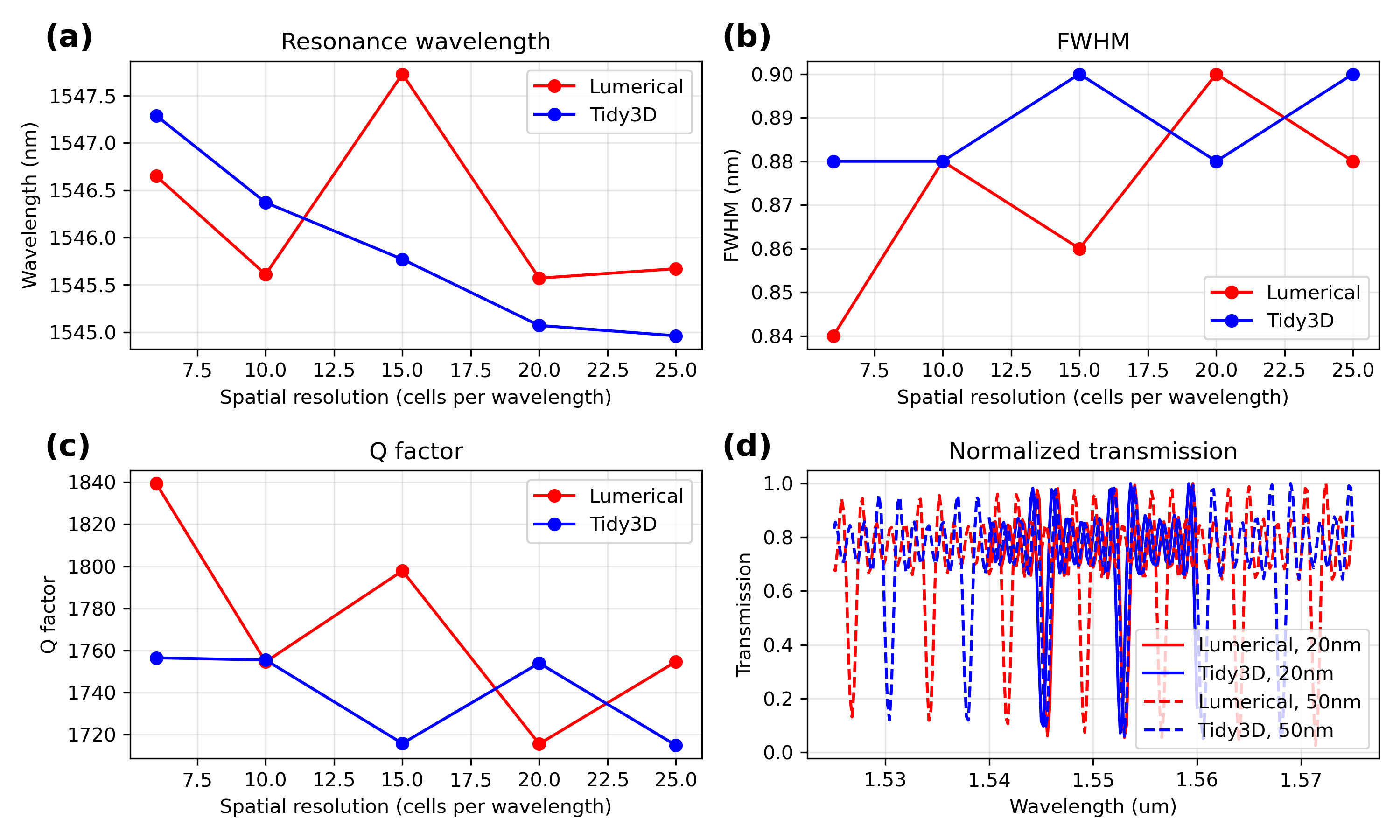}
	\caption{Simulation results of the ring resonator. (a) Lowest resonance wavelength in 20~nm source bandwidth for different spatial resolutions. (b) FWHM for different spatial resolutions. (c) Q-factor for different spatial resolutions. (d) Normalized Transmission spectra before interpolation with 20~nm and 50~nm source bandwidths at the spatial resolution of 20 cells per wavelength.}
	\label{fig:RING_results}
\end{figure}

The simulation results are presented in Fig.~\ref{fig:RING_results}.
The monitors in these simulation has a wavelength step size of 0.2~nm.
We then interpolate the transmission spectra from the output port with a step size of 0.02~nm for more accurate estimation of the lowest resonance wavelength~$\lambda_0$ in the simulation bandwidth and the corresponding full width at half maximum~(FWHM,~$\Delta\lambda$).
Also, we calculate the quality factor, namely the Q-factor, as following:
\begin{equation}
	Q = \frac{\lambda_0}{\Delta\lambda}.
\end{equation}
As presented in Fig.~\ref{fig:RING_results}(a), Lumerical FDTD gives a resonance wavelength around 1545.6~nm for resolutions higher than 10 cells per wavelength, except for 15 where an abrupt increase occurs.
Meanwhile, Tidy3D shows a gradual decrease in resonance wavelength and converges to around 1545.0~nm at 20 cells per wavelength.
The corresponding FWHM is included in Fig.~\ref{fig:RING_results}(b).
Lumerical FDTD shows an increase from 0.84~nm at 6 cells per wavelength to 0.88~nm at 25 cells per wavelength with fluctuations in between, while Tidy3D maintains in the range between 0.88~nm and 0.9~nm.
The difference between the two solvers reaches a stable value of 2$\%$ above 20 cells per wavelength.
Similar trends can be found for the Q-factor presented in Fig.~\ref{fig:RING_results}(c), where Lumerical FDTD generally decreases from 1839.4 to 1754.7 and Tidy3D oscillates between 1756.5 and 1714.9.
The inter-solver difference in Q-factor also reaches 2$\%$ for resolutions higher than 20 cells per wavelength. 

\begin{figure}[htbp]
	\centering
	\includegraphics[width=0.5\linewidth]{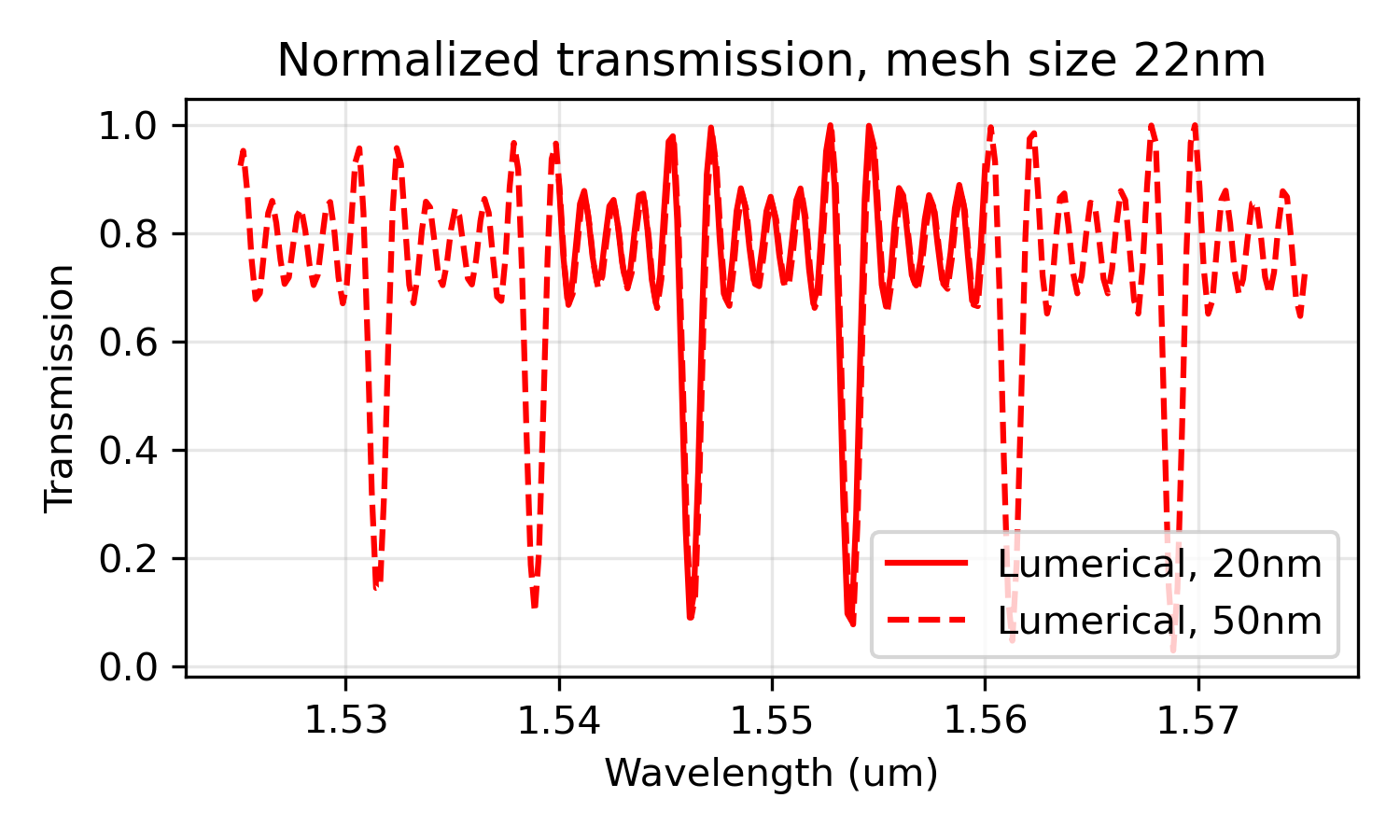}
	\caption{Normalized transmission spectra before interpolation with 20~nm and 50~nm source bandwidths computed with a mesh size of 22~nm in Lumerica FDTD}
	\label{fig:RING_mesh}
\end{figure}

We then evaluate the performance over different source bandwidths.
The transmission spectra from the two solvers for 20~nm and 50~nm bandwidth are shown in Fig.~\ref{fig:RING_results}(d).
When the source bandwidth is increased from 20~nm to 50~nm in Lumerical FDTD, the resonance wavelengths shift by 3.5~nm in Lumerical FDTD, which is about 47$\%$ of the free spectral range~(FSR) of around 7.4~nm.
In contrast, the resonance wavelengths shift up by 0.4~nm in Tidy3D, around 5$\%$ of the FSR of 7.6~nm.
Similar to the discussions in previous sections, it suggests a potential impact of mesh size variation induced by custom non-uniform mesh and changed simulation bandwidth.
This is proven by switching to a uniform mesh size of 22~nm over all three dimensions, where the difference is significantly reduced as shown in Fig.~\ref{fig:RING_mesh}.

\section{Discussion}

In the previous section, we present the simulation results of a selection of passive components on SOI platform sourced from GDSFactory generic PDK, including directional coupler, waveguide crossing, 2$\times$2 MMI, mode converter, a polarization splitter rotator~(PSR), and a ring resonator.
Through the simulation of directional coupler and waveguide crossing, we evaluate the propagation and coupling of fundamental modes.
The multimode behavior is examined by the simulation of MMI.
Moreover, the simulation of mode converter and PRS demonstrates higher-order mode coupling and polarization transitioning.
Lastly, we compare the resonance behavior of ring resonator between the two solvers.
The minimum feature size across these components is 200~nm, occurring in the MMI between the tapers and in the ring resonator between the ring and the straight waveguide.
In general, Tidy3D shows finer discretization given the same spatial resolution setting and elapses longer simulation time before reaching the auto-shutoff threshold.
The two solvers give similar runtime when run on advanced NVIDIA GPUs.
It needs to be noted that the latest version of Lumerical FDTD (R2.1) does not support frequency dependent mode profiles when solving on GPUs.
This could potentially impact broadband simulations.

The key results are summarized in Table~\ref{tab:summary}.
A spatial resolution of 15 cells per wavelength is sufficient for most components in the two solvers.
It corresponds to a mesh size of approximately 30~nm, as defined by the wavelength~(1550~nm) and the refractive index of Si.
The simulation of ring resonator is more demanding and requires a higher resolution of 20 cells per wavelength.
On one hand, in the simulation of waveguide crossing, Tidy3D reaches a stable output at a lower resolution~(10 cells per wavelength) than Lumerical FDTD.
On the other hand, in the simulation of mode converter and PSR, Lumerical FDTD achieves stable results at lower resolutions than Tidy3D.

\begin{table}[htbp]
	\small
	\caption{Key simulation results of the benchmark cases}
	\label{tab:summary}
	\centering
	\begin{tabular}{|c|c|c|c|}
		\hline
		Component & Metrics & Lumerical FDTD & Tidy3D \\
		\hline
		\multirow{4}{*}{\shortstack{diectional\\coupler}} & \makecell{Required resolution\\(cells per wavelength)} & 15 & 15 \\ \cline{2-4}
		& Bandwidth-induced variation & 3$\%$ & 0.2$\%$ \\ \cline{2-4}
		& Inter-solver discrepancy & \multicolumn{2}{c|}{3.7$\%$} \\ 
		\hline
		\multirow{4}{*}{\shortstack{waveguide\\crossing}} & \makecell{Required resolution\\(cells per wavelength)} & 15 & 10 \\ \cline{2-4}
		& Bandwidth-induced variation & 0.1$\%$ & 0.03$\%$ \\ \cline{2-4}
		& Inter-solver discrepancy & \multicolumn{2}{c|}{0.2$\%$} \\ 
		\hline
		\multirow{4}{*}{\shortstack{2$\times$2\\MMI}} & \makecell{Required resolution\\(cells per wavelength)} & 15 & 15 \\ \cline{2-4}
		& Bandwidth-induced variation & 0.05$\%$ & 0.05$\%$ \\ \cline{2-4}
		& Inter-solver discrepancy & \multicolumn{2}{c|}{0.5$\%$} \\ 
		\hline
		\multirow{4}{*}{\shortstack{mode\\converter}} & \makecell{Required resolution\\(cells per wavelength)} & 10 & 15 \\ \cline{2-4}
		& Bandwidth-induced variation & 9$\%$ & 3$\%$ \\ \cline{2-4}
		& Inter-solver discrepancy & \multicolumn{2}{c|}{16$\%$} \\ 
		\hline
		\multirow{4}{*}{PRS} & \makecell{Required resolution\\(cells per wavelength)} & 15 & 20 \\ \cline{2-4}
		& Bandwidth-induced variation & 6$\%$ & 0.1$\%$ \\ \cline{2-4}
		& Inter-solver discrepancy & \multicolumn{2}{c|}{2$\%$} \\ 
		\hline
		\multirow{4}{*}{\shortstack{ring\\resonator}} & \makecell{Required resolution\\(cells per wavelength)} & 20 & 20 \\ \cline{2-4}
		& Bandwidth-induced variation & 47$\%$ & 0.5$\%$ \\  \cline{2-4}
		& Inter-solver discrepancy & \multicolumn{2}{c|}{2$\%$} \\
		\hline
		
	\end{tabular}
\end{table}

The two solvers generally agree well in a moderate simulation bandwidth of 20~nm over all examined components.
The inter-solver discrepancy is quantified by examining the transmission to the target mode of each component, except for the ring resonator where we take the difference in FWHM and Q-factor.
The difference is negligible for waveguide crossing and 2$\times$2 MMI, and below 4$\%$ for directional coupler, PRS, and ring resonator.
A higher difference of 16$\%$ is observed in the simulation of mode converter, a more demanding task where both solvers show considerable fluctuations at mismatched positions.

A change in the source bandwidth affects the two solvers differently.
Tidy3D shows little variation as the bandwidth changes.
However, when the bandwidth is increased from 20~nm to 50~nm, Lumerical FDTD sometimes gives different results, especially for the more complex components like the mode converter, PRS, and ring resonator.
This variation can be reduced significantly by switching from custom non-uniform meshing to a uniform mesh.
The former depends on not only the specified number of cells per wavelength but also the shortest wavelength in the simulation bandwidth.
Note that the bandwidth-induced variation for ring resonator quantifies the shift of resonance wavelength as a percentage of FSR, while other components are evaluated based on the transmission to the target mode.

It worth noting that Tidy3D encountered the problem of diverging simulation in the simulations of directional coupler at higher resolutions~(20 and 25 cells per wavelength) and mode converter across all resolutions.
This is overcome by changing the boundary condition from PML to ``absorber''.
Boundary conditions do not need to be changed for our simulations in Lumerical FDTD.

\begin{figure}[htbp]
	\centering
	\includegraphics[width=0.5\linewidth]{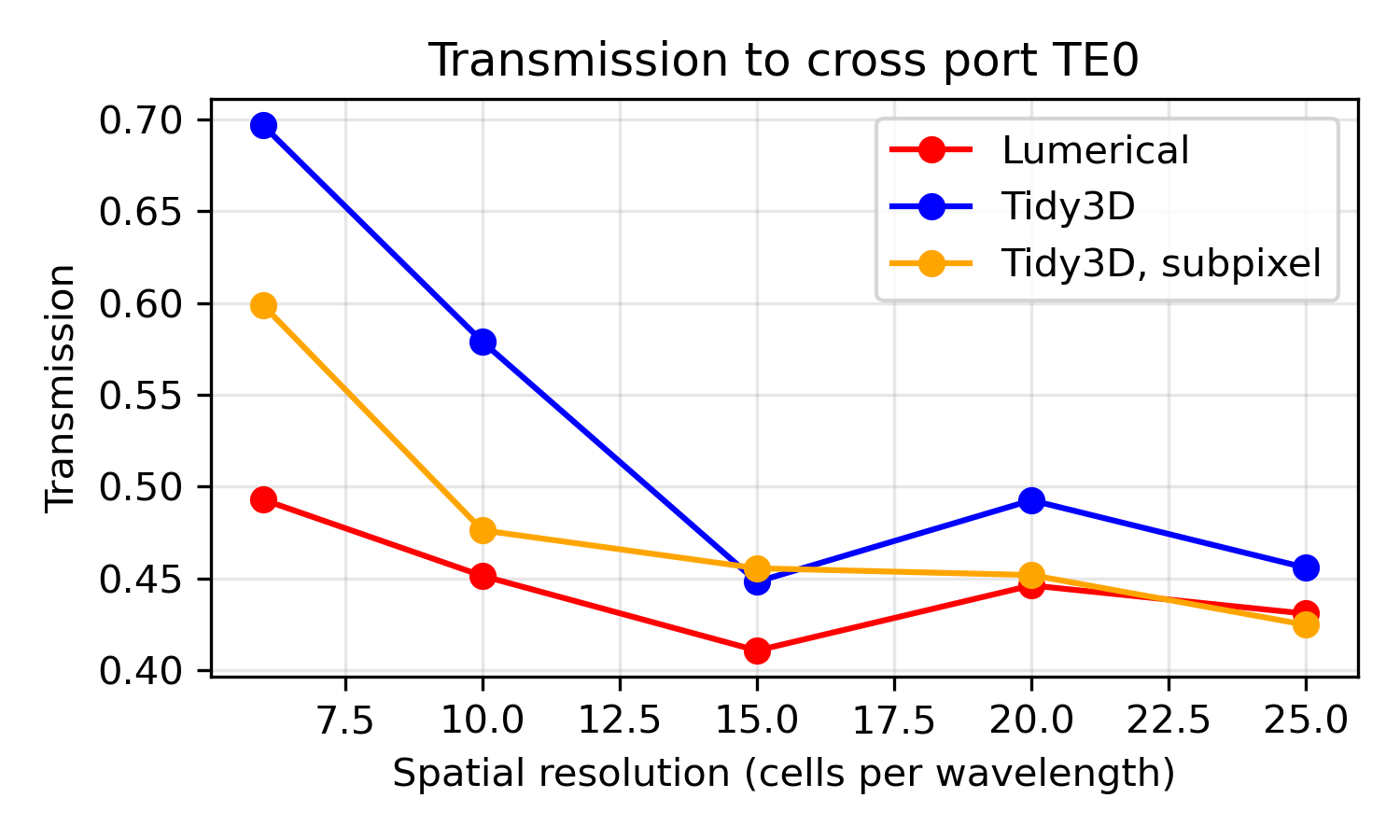}
	\caption{Simulation result of directional coupler using subpixel averaging in Tidy3D: transmission to the fundamental TE mode at the cross port at 1550~nm.}
	\label{fig:DC_subpixel}
\end{figure}

The Tidy3D simulations presented in the previous sections are executed using version 2.7.7.
During the course of our work, a newer version 2.8 has been released.
It was brought to our attention that a new scheme named subpixel averaging is added so that the solver can better account for geometric features smaller than the mesh~\cite{Jin2025}, functionally similar to CMT in Lumerical FDTD mentioned in Section 2.
We examined the simulations of the directional coupler again using version 2.8.5 and subpixel averaging.
As shown in Fig.~\ref{fig:DC_subpixel}, transmission to the target mode at 1550~nm converges to around 47$\%$ at a lower spatial resolution of 10 cells per wavelength.

Both Lumerical FDTD and Tidy3D are proven, reliable solvers for FDTD simulations.
Lumerical FDTD is a mature solver offering extensive features and configurable settings.
It can run efficiently on local and cloud-based hardware.
The software includes a comprehensive example library and a scripting language.
It also offers Python plugins (e.g. \texttt{lumapi}, \texttt{lumopt}).
It can also be used in combination with other multiphysics solvers and circuit simulator offered by Lumerical.
Tidy3D uses cloud-native GPU acceleration for rapid simulations, features a clear Python API and extensive examples, and is under active development to enhance its multiphysics and circuit simulation capabilities.

\section{Conclusions}

We presented a systematic comparison between Lumerical FDTD and Tidy3D across a range of passive components on SOI platforms representing different simulation scenarios.
Our study focused on quantitative metrics including simulation runtime, memory consumption, bandwidth sensitivity, and inter-solver discrepancy.
Under matched simulation conditions, the two solvers demonstrated excellent agreement in simulation results.
Additionally, for most components, both solvers showed minimal variation in the output when the simulation bandwidth is increased from 20~nm to 50~nm.
These results reinforce the reliability of both solvers.

\begin{backmatter}
    \bmsection{Funding}
    This work was supported by Natural Sciences and Engineering Research Council of Canada (NSERC, grant number RGPIN-2024-05029) and Mitacs through the Mitacs Elevate program.
    
    \bmsection{Acknowledgment}
    The authors thank Wesley Sacher for insightful discussions about the selection of devices and materials in this work.
    The authors also thank Flexcompute for providing information on the latest update and computation hardware of Tidy3D.
    Additionally, the authors thank Ansys for providing access to Ansys Cloud Burst Compute and discussion about definition of runtime.
    
    \bmsection{Disclosures}
    The authors declare no conflicts of interest.
    
    \bmsection{Data Availability}
    The Python code underlying the results presented in this paper are available in Ref.~\cite{Liu2025}.
    
\end{backmatter}
    
\bibliography{sample}    
    
\end{document}